\documentclass[12pt,preprint]{aastex}

 \usepackage{color}

\slugcomment{Accepted for publication in ApJ}

\shorttitle{Dust formation and Evolution in SN IIb}
\shortauthors{Nozawa et al.}

\begin{document}

\title{Formation and Evolution of Dust in Type II\lowercase{b} 
Supernova with \\ Application to the Cassiopeia A Supernova Remnant}

\author{
Takaya Nozawa,\altaffilmark{1} 
Takashi Kozasa,\altaffilmark{2} 
Nozomu Tominaga,\altaffilmark{3,1} 
Keiichi Maeda,\altaffilmark{1} \\ 
Hideyuki Umeda,\altaffilmark{4}
Ken'ichi Nomoto,\altaffilmark{1,4} and
Oliver Krause\altaffilmark{5}}

\altaffiltext{1}{Institute for the Physics and Mathematics of the
Universe, University of Tokyo, Kashiwa, Chiba 277-8568, Japan; 
tnozawa@mail.sci.hokudai.ac.jp}
\altaffiltext{2}{Department of Cosmosciences, Graduate School 
of Science, Hokkaido University, Sapporo 060-0810, Japan}
\altaffiltext{3}{Department of Physics, Faculty of Science and 
Engineering, Konan University, Okamoto, Kobe 658-8501, Japan}
\altaffiltext{4}{Department of Astronomy, School of Science, 
University of Tokyo, Bunkyo-ku, Tokyo 113-0033, Japan}
\altaffiltext{5}{Max-Planck-Institut f\"ur Astronomie, K\"onigstuhl 17, 
69117 Heidelberg, Germany}

\begin{abstract}
How much amount and what size of dust are formed in the ejecta of 
core-collapse supernovae (CCSNe) and are injected into the interstellar 
medium (ISM) depend on the type of CCSNe through the thickness of outer 
envelope.
Recently Cas A was identified as a Type IIb SN (SN IIb) that is 
characterized by the small-mass hydrogen envelope.
In order to clarify how the amount of dust formed in the ejecta and 
supplied into the ISM depends on the type of CCSNe, we investigate the 
formation of dust grains in the ejecta of a SN IIb and their evolution in 
the shocked gas in the SN remnant (SNR) by considering two sets of 
density structures (uniform and power-law profiles) for the circumstellar 
medium (CSM).
Based on these calculations, we also simulate the time evolution of 
thermal emission from the shock-heated dust in the SNR and compare the 
results with the observations of Cas A SNR.
We find that the total mass of dust formed in the ejecta of a SN IIb is 
as large as 0.167 $M_{\odot}$ but the average radius of dust is smaller 
than 0.01 $\mu$m  and is significantly different from those in SNe II-P 
with the massive hydrogen envelope; in the explosion with the small-mass 
hydrogen envelope, the expanding He core undergoes little deceleration, 
so that the gas density in the He core is too low for large-sized grains 
to form.
In addition, the low-mass hydrogen envelope of the SN IIb leads to the 
early arrival of the reverse shock at the dust-forming region.
If the CSM is more or less spherical, therefore, the newly formed small 
grains would be completely destroyed in the relatively dense shocked gas 
for the CSM hydrogen density of $n_{\rm H} > 0.1$ cm$^{-3}$ 
without being injected into the ISM.
However, the actual CSM is likely to be non-spherical, so that a part of 
dust grains could be ejected into the ISM without being shocked.
We demonstrate that the temporal evolution of the spectral energy 
distribution (SED) by thermal emission from dust is sensitive to the 
ambient gas density and structure that affects the passage of the reverse 
shock into the ejecta. Thus, the SED evolution well reflects the evolution 
of dust through erosion by sputtering and stochastic heating.
For Cas A, we consider the CSM produced by the steady mass loss of 
$\dot{M} \simeq 8 \times 10^{-5}$ $M_\odot$ yr$^{-1}$ during the 
supergiant phase.
Then we find that the observed infrared SED of Cas A is reasonably 
reproduced by thermal emission from the newly formed dust of 0.08 
$M_\odot$, which consists of the 0.008 $M_\odot$ shock-heated warm dust 
and 0.072 $M_\odot$ unshocked cold dust. 
\end{abstract}

\keywords{dust, extinction --- infrared: ISM --- ISM: supernova remnants 
--- shock waves --- supernovae: general --- supernovae: individual 
(Cassiopeia A)}

\section{Introduction}

Core-collapse supernovae (CCSNe) have been considered to be one of
important sources of interstellar dust.
They eject a large amount of heavy elements synthesized during their 
stellar evolution and explosive burning, and form dust grains inside the 
dense, metal-rich He core in the expanding ejecta.
Meanwhile, the reverse and forward shocks driven by the encounter of the 
ejecta materials with the surrounding medium propagate into the SN ejecta 
and the interstellar medium (ISM), respectively, producing the hot gas 
plasma within the supernova remnant (SNR). 
The dust grains formed in the He core, before being dispersed into the 
interstellar space, pass through the shock-heated gas and suffer from 
erosion and destruction by sputtering due to collisions with the 
energetic ions.
Since the efficiency of destruction of dust grains heavily depends on 
their composition and size, what composition and what size of dust 
condense in the ejecta should be clarified for understanding the 
properties and amount of dust finally injected from CCSNe into the ISM.

Observations of nearby CCSNe have reported increasing evidence for dust 
formation in the ejecta (see Kozasa et al.\ 2009 for review).
Nevertheless, the composition and size of newly formed dust have not been 
specified except for two cases; SN 2004et (Kotak et al.\ 2009) and SN 
2006jc (Sakon et al.\ 2009; Smith et al.\ 2008; Mattila et al.\ 2008). 
The masses of dust formed in the SN ejecta are estimated to be less than 
$10^{-3}$ $M_\odot$ from the observations by usually assuming that the 
ejecta is optically thin for thermal emission of dust.
However, newly formed dust may reside in optically thick clumps (e.g., 
Miekle et al.\ 2007).
Moreover, some species of dust grains can cool down quickly after the 
formation (Nozawa et al.\ 2008), so that near- to mid-infrared (IR) 
observations could miss the cold dust.
Hence, the derived mass of dust could be the lower limit in some cases, 
and the sophisticated radiative transfer calculations (e.g., Sugerman et 
al.\ 2006; Ercolano et al.\ 2007) referring to the observational data 
at UV/optical to far-IR wavelengths would be required to deduce the 
properties and mass of dust formed in the ejecta from observations.

Theoretical studies have shown that various species of dust grains with 
the radii of 0.001--1 $\mu$m and total mass of 0.1--1 $M_\odot$ can 
condense in the ejecta of Type II-P SNe (SNe II-P) retaining the massive 
hydrogen envelope (Kozasa et al.\ 1989, 1991; Todini \& Ferrara 2001; 
Nozawa et al.\ 2003).
Adopting the results of dust formation calculations, a semi-analytical 
model (Bianchi \& Schneider 2007) and a numerical simulation (Nozawa et 
al.\ 2007) investigated the survival of dust in the SNRs and found that 
dust mass injected into the uniform ISM is $\sim$0.01--0.8 $M_\odot$ 
for the ambient hydrogen density of $n_{\rm H,0} =$ 10--0.1 cm$^{-3}$.
This mass of dust supplied by SNe II-P is still large enough to explain 
a huge quantity of dust in excess of $10^8$ $M_\odot$ observed for the 
host galaxies of quasars at redshift $z > 5$ (Morgan \& Edmunds 2003; 
Maiolino et al.\ 2006; Dwek et al.\ 2007), though it is proposed that 
asymptotic giant branch stars can also contribute to dust enrichment 
even in such an early epoch (Valiante et al.\ 2009).

Recently Kozasa et al.\ (2009) have suggested that how much amount and 
what size of dust are formed in the ejecta and are injected into the ISM 
depend on the type of CCSNe, reflecting the thickness of the H-rich 
envelope.
The less massive outer envelope leads to extremely high expansion 
velocity of the He core unless the kinetic energy of explosion is very 
low, and makes the gas density in the expanding ejecta of the He core too 
low for large-sized grains to be produced.
Furthermore, the low-mass outer envelope can allow the early arrival of 
the reverse shock at the dust-forming region and result in efficient 
destruction of newly formed dust in the shock-heated gas with high
density.
Because a significant fraction of CCSNe ($\sim$0.3) explode as 
hydrogen/helium-poor CCSNe such as SNe Ib/Ic and SNe IIb (Prieto et al.\ 
2008; Smartt et al.\ 2009; Boissier \& Prantzos 2009), it is essential to 
investigate how the formation and destruction processes of dust depend on 
the type of CCSNe, in order to reveal the role of CCSNe as sources of 
interstellar dust.

It should be mentioned here that dust grains embedded in the postshock 
gas are eroded by sputtering and are heated up by collisions with the 
hot gas, radiating their thermal emission at IR wavelengths.
Since the destruction efficiency and heating rate of dust are sensitive 
to grain size as well as gas density, the IR spectral energy distributions 
(SEDs) by thermal emission from dust in SNRs can be a diagnostic of the 
size and temperature of dust, and therefore provide key information on the 
nature of dust formed in the ejecta, density structure in the ambient 
medium, and destruction and heating processes of dust.
In particular, young SNRs that originated from SNe with the low-mass 
envelope are ideal ones to compare the observed IR SEDs with calculated 
ones; the reverse shock can quickly arrive at the He core and heat up newly 
formed dust before the forward shock sweeps up a considerable amount of 
circumstellar/interstellar dust, so that we can mainly see the IR emission 
from dust grains condensed in the ejecta.
In this sense, Cassiopeia A (Cas A) SNR is one of the best targets for 
this study because the type of the Cas A SN is recently identified as Type 
IIb by a striking similarity between the optical spectrum of scattered 
light echo of Cas A SN and the time-averaged spectrum of SN IIb 1993J 
(Krause et al.\ 2008), and the presence of dust freshly formed in the 
ejecta has been confirmed by IR observations (e.g., Rho et al.\ 2008).

In this paper we explore the formation of dust in a SN IIb and the 
evolution of the dust inside the SNR, with the aim of revealing the 
composition, size, and mass of dust ejected from hydrogen-deficient CCSNe.
In addition, the time evolution of thermal emission from dust within 
SNRs is calculated and is compared with the observations of Cas A to 
discuss the composition and amount of dust.
In \S~2 we perform the calculation of dust formation in the ejecta of a
SN IIb whose progenitor has lost most of the hydrogen envelope during the 
evolution (Nomoto et al.\ 1993; Podsiadlowski et al.\ 1993).
In \S~3 we investigate the evolution of the newly formed dust in the gas 
heated by the reverse and forward shocks and its dependence on the 
ambient gas density and structure.
Then, following the calculations of dust evolution in SNRs, we present the
temporal evolution of IR emission from collisionally heated dust in the 
hot gas in \S~4, and compare the results with the observations of Cas A 
in \S~5.
We summarize our results in \S~6.

\section{Formation of Dust in the Ejecta of SN II\lowercase{b}}

In the ejecta of SNe, dust formation takes place within the He core where 
the densities of condensible elements are high enough to advance the 
formation and growth of seed nuclei through collisions with the relevant 
gas species.
Formation processes of dust grains are influenced by the time evolution 
of the density and temperature of the gas as well as the elemental 
composition in the ejecta (Kozasa et al.\ 1989).
In this section, we describe the model of the dust formation calculation
for a SN IIb and present the results of the calculations.

\subsection{Model of SN IIb and Calculation of Dust Formation}

The calculation of dust formation is carried out by applying the theory 
of non-steady state nucleation and grain growth developed by Nozawa et 
al.\ (2003).
The theory takes into account chemical reactions at condensation,
considering that the kinetics of nucleation and grain growth is 
controlled by the key species defined as the gas species with the least 
collisional frequency among the reactants (Kozasa \& Hasegawa 1987; 
Hasegawa \& Kozasa 1988).
Given the elemental composition of the gas in the ejecta as well as the
time evolution of the gas density and temperature, the theory allows us 
to determine the condensation time, chemical composition, size 
distribution, and mass of dust that condenses in the ejecta (see Nozawa 
et al.\ 2003 for details).

In the calculation of dust formation, we adopt the model of the SN IIb 
with the zero-age main-sequence mass of 18 $M_\odot$ and metallicity of 
$Z=0.02$. 
The ejecta mass is $M_{\rm eje} = 2.94$ $M_\odot$, the mass of hydrogen 
envelope is $M_{\rm env} = 0.08$ $M_\odot$, the kinetic energy of 
explosion is $E_{\rm kin}=10^{51}$ ergs, and the synthesized $^{56}$Ni 
mass is $M(^{56}{\rm Ni})=$ 0.07 $M_\odot$.
These model parameters are invoked to reproduce the light curve of SN 
1993J by referring to the SN model in Shigeyama et al.\ (1994).
Note that the mass of ejecta in this model is in good agreement with the 
ejecta mass of Cas A ($=$ 2--4 $M_\odot$) estimated from the studies of 
shock dynamics (Borkowski et al.\ 1996; Chevalier \& Oishi 2003; Laming 
\& Hwang 2003) and from X-ray observations (Vink et al.\ 1996; Favata et 
al.\ 1997; Willingale et al.\ 2003).

In Figure 1 we show the elemental composition in the ejecta of the SN 
IIb model at 600 days after the explosion, taking the decay of 
radioactive elements into account.
In Table 1 we also present the mass of the main refractory species 
$i$ in the ejecta.
We assume that the elemental distribution in the ejecta remains an 
onion-like structure without any mixing, which is consistent with the 
theoretical model of mixing in SN 1993J;
Iwamoto et al.\ (1997) have shown with the 2D simulation that the 
deceleration of the He core by the thin H-rich envelope is so weak that 
the growth of the Rayleigh-Taylor instability is too slow to cause 
substantial mixing in the ejecta, in contrast to SN 1987A which has a 
thick H-rich envelope (e.g., Hachisu et al.\ 1990, 1992).
Although the jet formation in the explosion suggested from the 
observations of Cas A can lead to macroscopic mixing (e.g., Tominaga 
2009), a simple argument of molecular diffusion has shown that mixing 
at the atomic level is extremely ineffective before the condensation of 
dust grains in the ejecta (Deneault et al.\ 2003).
Indeed, the observations of Cas A at various wavelengths have suggested 
that the mixing in the ejecta should not be at atomic level but at the 
macroscopic level 
(e.g., Willingale et al.\ 2002; Fesen et al.\ 2006b; Ennis et al.\ 2006).
From Figure 1, we can see the layered structure of elemental composition 
with the very thin hydrogen envelope at the outermost ejecta, which is 
followed by C-rich (C/O $>$ 1) He layer, O-rich layer, Si-S-Fe layer, and 
Fe-Ni layer inwardly.

The distributions of gas temperature and density in the ejecta of the 
SN IIb at 300 days and 600 days are shown in Figures 2{\it a} and 
2{\it b}, compared with those for the SN II-P model of
$M_{\rm eje} = 17$ $M_\odot$ and $M_{\rm env}$ =  13.2 $M_\odot$ with
the same $E_{\rm kin}$ and $M(^{56}{\rm Ni})$ as those of the SN IIb 
model (Umeda \& Nomoto 2002).
The gas density scales as $t^{-3}$ with the time $t$ after the explosion,
and the time evolution of gas temperature is mainly regulated by the 
energy deposited from radioactive elements (i.e.\ $M(^{56}{\rm Ni})$).
It should be noted that the gas density in the He core of the SN IIb is 
about three orders of magnitude lower than that in the SN II-P.
The reason for this is that, since almost all of the explosion energy 
are deposited into the He core without the thick hydrogen envelope, the 
resulting expansion velocity of the gas in the He core of the SN IIb is 
much higher than that of the SN II-P, as shown in Figure 2{\it c}. 

Based on the hydrodynamic and nucleosynthesis models for the SN IIb, we 
calculate dust formation.
Dust formation proceeds through the formation and growth of seed nuclei 
as the gas cools down.
The number of seed nuclei is evaluated by calculating the non-steady state 
nucleation rate (Eq (5) in Nozawa et al.\ 2003) being coupled with the 
time evolution of the gas temperature and the number density of the key 
species.
In the calculations, it is assumed that without coagulation the seed 
nuclei grow homogeneously at a rate controlled by the collisions of the
key species with the sticking probability of unity.
We also assume that the formation of CO and SiO molecules is complete; 
all C (Si) atoms are locked in CO (SiO) molecules in the case where 
the number density of C (Si) atoms is lower than that of O atoms.
Under this assumption, C-bearing grains can condense only in the He layer 
with C/O $>$ 1.
Depletion or rarefaction of the key species due to the grain growth or
the expansion of the ejecta, respectively, causes the process of 
nucleation and grain growth to cease in less than a few tens of day since 
the onset of nucleation.

\subsection{Results of Dust Formation Calculation}
 
The simulation results of dust formation in the SN IIb are illustrated 
in Figures 3{\it a} and 3{\it b}, where the mass distribution and 
average radius of each dust species formed in the ejecta are given as a 
function of enclosed mass $M_r$, respectively. 
Without any mixing of elements, different species of dust condense 
according to the elemental composition of the gas in each layer:
C grains in the He layer, silicates (MgSiO$_3$, Mg$_2$SiO$_4$, and
SiO$_2$) and oxides (Al$_2$O$_3$, and MgO) in the O-rich layer, and 
FeS and Si grains in Si-S-Fe layer.
The typical times (gas temperatures) at which dust grains condense in 
the ejecta are 300--350 days (1400--1700 K) for C grains, 350--500 
days (1100--1400 K) for silicate and oxide grains, and 600--700 days 
(800--900 K) for Si and FeS grains.
In the innermost Fe-Ni layer, neither Fe nor Ni grains can be formed 
because the gas density becomes too low before the gas cools down to 
the temperatures ($\la$800 K) at which these grains can condense.
Fe atoms residing in both the O-rich and He layers are not abundant
enough to be locked in grains as Fe-bearing dust (see Fig.\ 1).
Except for Fe grains, the major species of dust formed in the SN IIb 
are the same as those for the unmixed cases in SNe II-P with $Z = 0$ in 
Nozawa et al.\ (2003).

The mass of each dust species formed in the ejecta of SN IIb is presented 
in Table 2, together with the ratio of its mass to the total dust mass
(see Table 1 for the mass and mass fraction of refractory elements locked 
in dust grains).
The total mass of newly formed dust is 0.167 $M_\odot$, which is more 
than two orders of magnitude larger than the dust mass estimated from a 
handful of observations of nearby dust-forming SNe.
C grains are the most dominant species in mass, with the mass of 0.07 
$M_\odot$, and the bulk of them is in the innermost He layer at 
$M_r =$ 2.7--2.9 $M_\odot$ (see Fig.\ 3{\it a}), where the mass fraction 
of available C atoms locked in C grains is $f_{\rm C} =$ 0.4--1.
The masses of MgSiO$_3$, Mg$_2$SiO$_4$, and SiO$_2$ are 0.055 $M_\odot$, 
0.017 $M_\odot$, and 0.016 $M_\odot$, respectively, and these silicate 
grains account for more than 50 \% of the total dust mass.
The mass distributions of silicate grains are influenced by the abundance
of SiO molecules (namely Si atoms, see Fig.\ 1), and fairly concentrate 
at $M_r =$ 1.7--1.9 $M_\odot$ with the mass fraction of the preexisting 
SiO molecules locked in silicate grains $f_{\rm Si} =$ 0.4--0.8.
The condensation efficiency, defined as the ratio of the total dust mass 
to the total metal mass, is $\simeq$ 0.13 for the SN IIb and lies in the 
range of those estimated for SNe II-P (0.05--0.25, Nozawa et al.\ 2003).

It should be emphasized here that, as can be seen from Figure 3{\it b}, 
the average radii of all grain species formed in the SN IIb are below 
0.01 $\mu$m and are much smaller than those ($a_{\rm ave} > 0.01$ $\mu$m) 
of dust grains formed in SNe II-P (Nozawa et al.\ 2003).
This difference is caused by the difference in the gas density within 
the expanding ejecta of the He core.
In general, a larger supercooling necessary for dust grains to condense
in a gas with lower density leads to the formation of a large number of
seed nuclei and results in smaller average radii even if all available 
gaseous atoms are used up for the growth of seed nuclei ($f_j = 1$, 
where $f_j$ is the mass fraction of key species $j$ locked in a given
grain species).
As was discussed in Kozasa et al.\ (1989), in the condition that the 
number density of key species is so low that the collision timescale 
of key species is longer than the expansion timescale of the gas,
neither a significant number of seed nuclei can condense nor nucleated 
grains can grow to sufficiently large radii via accretion onto the 
surface of the grains.

\subsection{Dependence of Grain Radius on SN Parameters}

Here we briefly discuss how our results on the dust formation are 
affected by the model parameters of SN explosion, in particular focusing 
on the kinetic energy of the explosion $E_{\rm kin}$.
A lower $E_{\rm kin}$ leads to a higher He core density, thus making it 
possible for more dust grains with larger radii to condense if the time 
evolution of the gas temperature could not be affected by the explosion 
energy.
At a fixed time $t_1$, the gas density is scaled by the expansion velocity
$v_{\rm eje}$ as $\rho_{\rm eje}(t_1) \propto v_{\rm eje}^{-3}$ for a given 
mass and density structure of the ejecta.
Since $v_{\rm eje} \propto (E_{\rm kin}/M_{\rm eje})^{1/2}$ and roughly 
$a_{\rm ave} \propto \rho_{\rm eje}^{1/3}$ (Nozawa et al.\ 2003), we find
$a_{\rm ave} \propto 1/v_{\rm eje} \propto (M_{\rm eje}/E_{\rm kin})^{1/2}$.
Thus, even for the extremely low explosion energy of 
$E_{\rm kin} = 10^{50}$ ergs, the increase in the average grain radii is 
as small as a factor of $\simeq$ 3.
In addition, if we consider $E_{\rm kin}=$ (1--4)$\times 10^{51}$ ergs 
estimated for Cas A (Willingale et al.\ 2003; Chevalier \& Oishi 2003; 
Laming \& Hwang 2003; Young et al.\ 2006), grain radii only reduce by a 
factor of $\la 2$, compared with those given in Figure 3{\it b}.

In reality, however, the assessed variation of the grain radius is 
suppressed by the dependence of the time evolution of gas temperature on 
$E_{\rm kin}$; 
with the same amount of $^{56}$Ni in the ejecta, the lower (higher) 
explosion energy can cause the gas temperature to decrease more slowly 
(more quickly) because of more (less) efficient energy deposition in the 
ejecta with the higher (lower) gas density.
This leads to the later (earlier) condensation of dust, which compensates 
for the increase (decrease) in the gas density.
Consequently, the condensation of dust occurs at the similar gas density,
and not only the radius but also the mass of dust formed in the ejecta is 
almost the same, regardless of the explosion energy (see Appendix A in 
Nozawa et al.\ 2003), as long as the range of variation of $E_{\rm kin}$
is not extremely large.
The qualitative analysis indicates that grain radius is insensitive to 
the explosion energy, and that our results on dust formation are
robust for the expected variation of model parameters for Cas A SN.

On the other hand, it should be noted that grain radius is greatly 
affected by the structure of the progenitor star resulting from the 
presence or absence of the massive outer envelope; 
as seen from Figure 2c, the He-core velocity of the SNe IIb with the 
thin hydrogen envelope is 5--10 times higher than that of the SN II-P 
with the thick envelope.
The higher velocity of the SN IIb leads to a few orders of magnitude 
lower gas density and thus produces about one order of magnitude smaller 
dust grains than the SN II-P.
Nozawa et al.\ (2008) have also pointed out the effect of the lack of the 
outer envelope on the size of newly formed dust in Type Ib SN 2006jc, 
applying the SN model by Tominaga et al.\ (2008).
They found that too low gas density in the He core, which is caused by 
the absence of the outer envelope,
prevents dust grains with the radii larger than 0.01 $\mu$m from forming 
within the He core.
These results indicate that the radii of dust grains formed in SNe IIb 
and SNe Ib/Ic are to be generally smaller than those in SNe II-P with 
the thick outer envelope.

Our results demonstrate that the composition and mass of dust grains 
formed in the ejecta are almost independent of the type of CCSNe and 
the initial metallicity of the progenitors because the distribution and 
yield of metals in the He core are similar.
On the other hand, their radii strongly depend on the type of CCSNe; 
SNe with the lower-mass envelope produce much smaller dust grains 
($a_{\rm ave} < 0.01$ $\mu$m) than SNe II-P with the much more massive 
envelope.
As shown in the next section, small sizes of newly formed dust grains 
will determine their subsequent fates within the SNR.

\section{Evolution of Dust in Type II\lowercase{b} SNR}

Dust grains formed in the He core initially comove with the gas without 
suffering from any processings.
Upon encountering the reverse shock, the gas is quickly decelerated, 
while the dust grains decouple from the gas with high velocities 
relative to the gas, and intrude into the hot plasma region created by 
the passage of the reverse and forward shocks.
These shocked dust grains are decelerated by the drag force, eroded by 
thermal and/or kinetic sputtering, and heated up by collisions with the 
hot gas.
The motion, destruction, and heating of dust depend on not only the 
properties of dust but also the density and temperature in the postshock 
gas.
In this section, applying the result of the dust formation calculation, 
we examine the evolution of the dust in the hot gas within SNRs and its
dependence on the gas density and structure in the CSM.

\subsection{Model of Calculations for Destruction and Motion of Dust 
in SNRs}

The calculations of dust evolution within SNRs follow the method 
described by Nozawa et al.\ (2007); 
dust grains are treated as test particles, and the motion and destruction 
of dust are pursued by calculating self-consistently the erosion due to
sputtering and deceleration due to the gas drag together with the time 
evolution of gas temperature and density in the spherical symmetry shock.
We make use of the size distribution and spatial distribution of each dust 
species obtained in the last section as the initial conditions, and adopt 
the model of the SN IIb used in \S~2 as those for the velocity and density 
structure in the ejecta.

For the structure of circumstellar medium (CSM), which affects the 
evolution of SNRs and thus has critical impacts on the evolution of dust, 
we consider two sets of density profile. 
One is the uniform CSM with the hydrogen number densities of 
$n_{\rm H,0}$= 0.1, 1, and 10 cm$^{-3}$, assuming that the gas in the CSM 
has the solar composition and are fully ionized.
For the other CSM density profile, we consider the density structure 
produced by the steady stellar wind during the evolution of the progenitor;
the density profile is represented in terms of hydrogen number density as
\begin{eqnarray}
n_{\rm H}(r) &=& n_{\rm H,1} \left( \frac{r}{r_1} \right)^{-2} 
~ {\rm cm}^{-3} ~~ {\rm for} ~ r < r_2
\nonumber \\
 &=& 1 ~ {\rm cm}^{-3} ~~~~~~~~~~~~~~~~~~ {\rm for} ~ r \ge r_2,
\end{eqnarray}
where $r_1 = 1.2 \times 10^{18}$ cm is the radius of the freely expanding 
outermost ejecta at $t = 10$ yr after the explosion in the present model, 
and $r_2$ is the radius such that 
$n_{\rm H,1} (r_2/r_1)^{-2} =1$ cm$^{-3}$.
The coefficient $n_{\rm H,1}$ is given as
\begin{eqnarray}
n_{\rm H,1} = n_{\rm H}(r_1) = 
\frac{\dot{M} A_{\rm H}}{4 \pi r_1^2 v_w \mu m_{\rm H}},
\end{eqnarray}
where $\dot{M}$ is the mass-loss rate, $v_w$ is the stellar wind velocity, 
$\mu$ is the mean molecular weight, and $m_{\rm H}$ and $A_{\rm H}$ are 
the mass and the number abundance of hydrogen atom, respectively. 
In the calculations, we consider three cases of $n_{\rm H,1} = 30, 120$, 
and 200 cm$^{-3}$; for example, the case of $n_{\rm H,1}$= 30 cm$^{-3}$ 
with $v_w =$ 10 km s$^{-1}$ corresponds to $\dot{M} = 2 \times 10^{-5}$ 
$M_\odot$ yr$^{-1}$, which is the mass-loss rate of Cas A's progenitor 
suggested from the analysis of shock dynamics by Chevalier \& Oishi (2003).

The CSM models used in the calculations are listed in Table 3, and the 
uniform CSM and the stellar wind CSM ($\rho \propto r^{-2}$) are referred 
to as the model A and B, respectively, with the value of $n_{\rm H,0}$ or 
$n_{\rm H,1}$ attached for each case.
In the calculations, the spatial bin for hydrodynamic calculation is 
linearly divided into 2510 bins, and the size bin of dust is 
logarithmically divided into 80 bins for the range of 1 \AA~to 1 $\mu$m.
We assume that the gas in the CSM has a temperature of $10^4$ K and that 
the ejecta collide with the CSM at $t = 10$ yr after the SN explosion.
We have confirmed that the moderate changes of the CSM gas temperature 
($\sim$10$^3$--10$^5$ K) do not influence the results of calculations.
The calculations are performed up to $t = 10^5$--10$^6$ yr, depending on 
the CSM density.

\subsection{Results of Dust Evolution Calculation}

\subsubsection{Penetration of the Reverse Shock into the Ejecta}

We first describe how the penetration of the reverse shock depends on 
the gas density in the CSM.
Figure 4 exhibits the time evolution of radius and velocity of the 
reverse shock for the uniform CSM models A0.1, A1, and A10 with
$n_{\rm H,0} =$ 0.1, 1, and 10 cm$^{-3}$, respectively.
In the upper panel, we also show the trajectory of the boundary 
between the He core and the hydrogen envelope, for a reference, assuming
that the boundary is not decelerated even after the collision with the 
reverse shock.

The interaction of the SN ejecta with the ambient medium generates the 
forward shock running into the CSM to compress the ambient medium.
The shocked hot ambient gas thrusts the ejecta in turn, driving the 
reverse shock that propagates into the ejecta but initially expands 
outward in the Eulerian coordinate.
For the higher CSM density, the forward shock acquires a lower velocity, 
and the denser shocked gas is produced behind the forward shock. 
Accordingly, the reverse shock travels through the ejecta with a smaller 
outward velocity so that the unshocked ejecta can catch up with the 
reverse shock at earlier times (see Figure 4).
Thus, the time at which the reverse shock collides with the He core 
is earlier for the higher gas density in the CSM: $t_{\rm coll} =$ 
355 yr, 155 yr, and 75 yr for the model A0.1, A1, and A10, respectively.
For the stellar wind CSM, the collision times are earlier than for the 
uniform CSM models because of higher ambient density at the time of the 
ejecta-CSM interaction: $t_{\rm coll} =$ 63 yr, 48 yr, and 45 yr for the 
model B30, B120, and B200 with $n_{\rm H,1} =$ 30, 120, and 200 cm$^{-3}$ 
(see Table 3).

Note that these collision times for the remnant of Type IIb SN (hereafter 
Type IIb SNR) are a few tens times earlier than those for Type II-P SNRs, 
for which $t_{\rm coll} =$ (1.5--10) $\times 10^3$ yr 
for the CSM densities of $n_{\rm H,0} =$ 10--0.1 cm$^{-3}$ (Nozawa et 
al.\ 2007).
The earlier arrival of the reverse shock at the He core in Type IIb SNR 
is due to its much thinner hydrogen envelope than that of Type II-P SNR 
and the resulting higher expansion velocity of the He core (see Figure 
2{\it c}).
As shown below, the thickness of the hydrogen envelope has great 
effects on the evolution of dust in SNRs.

\subsubsection{Destruction of dust for the uniform CSM}

Figure 5 shows the time evolution of gas density ({\it upper panel}) and 
temperature ({\it lower panel}) within Type IIb SNR for the model A10 
(uniform CSM with $n_{\rm H,0}$= 10 cm$^{-3}$).
The gas densities behind the reverse and forward shocks, whose positions 
are depicted by the downward- and upward-pointing arrows in the lower 
panel, respectively, increase by a factor of about 4, compared with the 
unshocked ones.
The temperature of the gas swept up by both the shocks rises up to 
$>$10$^7$ K and remains over 10$^6$ K by $t \ga 10^5$ yr. 
Dust grains staying in this hot gas are subject to the destruction by 
sputtering.

Figure 6 presents the time evolution of positions ({\it upper panel})
and radii ({\it lower panel}) of dust grains with the initial radii of 
0.001 $\mu$m (for Al$_2$O$_3$ and FeS) and 0.003 $\mu$m (for the other 
grains) for the model A10, together with the trajectories of the reverse
and forward shocks in the upper panel.
In this model, the reverse shock encounters the He layer at $t = 75$ yr, 
the O-rich layer at $t = 270$ yr, and the Si-rich layer at $t = 700$ yr. 
Once newly formed dust grains intrude 
into the shocked gas, they are efficiently decelerated by the gas drag 
and are quickly trapped in the hot plasma. 
These grains are completely destroyed and are not injected into the ISM.
Actually all newly formed grains are destroyed in the postshock gas for 
most of the models considered in this paper.
One of the reasons for this is that the radii of newly formed grains are 
considerably small;
since the deceleration rate of dust is inversely proportional to the grain 
radius (e.g.,\ Nozawa et al.\ 2007), dust grains with the radii of $<$0.01 
$\mu$m are easily captured by the shock-heated gas and are destroyed by 
thermal sputtering.
Another reason is that the reverse shock can reach the He core more than 
20 times earlier for SN IIb than for SNe II-P as mentioned above;
although the initial gas density in the ejecta of the SN IIb is three 
orders of magnitude lower than those in SNe II-P (see Fig.\ 2b), 
at such an early time, the gas density in the He core of the SN IIb is 
still higher by a factor of $\sim$10 than those in SNe II-P at several
thousands years.
This leads to the frequent collisions between dust and gas and causes the 
efficient deceleration and destruction of dust.

In Figure 7{\it a} we compare the time evolution of the total dust mass 
for the uniform CSM with $n_{\rm H,0}$= 0.1, 1, and 10 cm$^{-3}$. 
We can see that dust grains are destroyed more efficiently and quickly 
for higher CSM density.
This is because the increase of the gas density in the CSM leads to the 
earlier arrival of the reverse shock at the dust-forming region, which 
keeps the gas density in the He core high enough for dust grains to be 
eroded by sputtering effectively.
For $n_{\rm H,0}$= 0.1 cm$^{-3}$, C grains of $\sim$10$^{-4}$ $M_\odot$
can survive at $t = 10^6$ yr, while all dust grains are destroyed within 
$t = 4 \times 10^5$ yr for $n_{\rm H,0}$= 1 cm$^{-3}$ and 
$t = 5 \times 10^3$ yr for $n_{\rm H,0}$= 10 cm$^{-3}$.

\subsubsection{Destruction of dust for the stellar wind CSM}

Figure 7{\it b} shows the evolution of the total dust mass in SNRs for 
the stellar wind CSM ($\rho \propto r^{-2}$) models.
For these models, dust grains are completely destroyed before
$t = 10^5$ yr.
However, the evolution of dust in SNRs shows the different behavior as 
shown in Figures 8{\it a} and 8{\it b}, where the time evolutions of 
mass of each dust species for the model B30 and B120 are presented.
For the model B30 with $n_{\rm H,1}$= 30 cm$^{-3}$, a small amount of 
C grains survive after $t = 10^4$ yr, whereas the other grain species 
are completely destroyed before $t \sim 10^4$ yr.
In the case of the model B120 with the CSM gas density four times higher 
than that for B30, C grains are destroyed quickly after the collision 
with the reverse shock, whereas the masses of some dust species such as 
SiO$_2$ and Si grains decrease more slowly than those for B30.

In order to explain these different behaviors of dust evolution, we show 
the time evolutions of density structure within SNRs for B30 and B120 in 
Figure 9{\it a} and 9{\it b}, respectively.
Since the deceleration and erosion rates are dependent on the number 
density of ion, the time evolutions of the total ion number density, 
$n$, are given in the lower panel in Figure 9, along with the positions 
of C, SiO$_2$, and Si grains with the initial radii of 0.003 $\mu$m.
We also present the time evolutions of positions and radii of these C, 
SiO$_2$, and Si grains in Figure 10. 

For the model B30 (Figs.\ 9{\it a} and 10{\it a}), the SNR expands into 
the CSM with the density profile of $\rho \propto r^{-2}$ during only 
340 yr, and then propagates into the ISM with the constant density of 
$n_{\rm H} = 1$ cm$^{-3}$. 
Thus, one might guess that the evolution of dust in the SNR may be 
quite similar to that for the model A1 with the uniform CSM of 
$n_{\rm H,0} = 1$ cm$^{-3}$.
However, as presented in Figure 7, dust grains in the ejecta are 
completely destroyed about 10 times earlier for B30 than for A1, because
the gas density behind the reverse shock for B30 is about 8 times higher 
than for A1 at $t \la 340$ yr due to the fast passage of the reverse 
shock into the ejecta (see Table 3).
In addition, the density of gas in the region swept up by the reverse 
shock at the initial phase remains to be quite high ($n \ga 0.3$ 
cm$^{-3}$) even after $t = 3000$ yr (see Fig.\ 9{\it a}; hereafter, 
such a region is referred to as the dense gas region).
As can be seen in Fig.\ 10{\rm a}, dust grains are trapped in the dense 
gas region shortly after their collisions with the reverse shock and 
are completely destroyed by thermal sputtering.
C grains formed in the inner C-rich region of the He layer collide with 
the reverse shock at an early epoch ($t \simeq 330$ yr) and are eroded by 
sputtering in the denser gas than those for SiO$_2$ and Si grains 
formed inside the O-rich layer.
However, C grains can survive for a relatively long period, compared
with SiO$_2$ and Si grains (see Fig.\ 8{\it a}), because the erosion
rate of C grains is smaller than the other grain species by a factor 
of about 5 (Nozawa et al.\ 2006).

In the case of the model B120 (Figs.\ 9{\it b} and 10{\it b}), the 
reverse shock more quickly penetrates into the ejecta than B30, which 
results in the extended shocked region between the reverse shock front 
and the dense gas region (see Fig.\ 9{\it b}).
C grains collide with the reverse shock at $t \simeq 140$ yr and are 
immediately destroyed in the shocked gas with $(330/140)^3 \simeq 13$
times higher density than that for B30. 
SiO$_2$ and Si grains, which encounter the reverse shock at 
$t \simeq 400$ yr and $t \simeq 700$ yr, respectively, intrude into 
the 30--40 times denser postshock gas than the cases for B30.
However, the gas density is not high enough for the SiO$_2$ and Si 
grains to be trapped and completely destroyed.
Instead, upon collisions with the reverse shock, these grains are 
greatly decelerated by the gas behind the reverse shock, which prevents
them from attaining the dense gas region in a short time;
SiO$_2$ grains are finally trapped in the extended shocked region and 
are destroyed without arriving at the dense gas region.
Si grains are eroded slowly passing through the extended shocked 
region until they can reach the dense gas region at $10^4$ yr.

For the model B200, the reverse shock can propagate into the ejecta 
even earlier than B120.
Hence, the gas density behind the reverse shock is so high that all grain 
species are trapped and are quickly destroyed once they are swept up by 
the reverse shock. 
These results demonstrate that the evolution of dust within SNRs strongly 
depends on the gas density and structure in the CSM, as well as on the 
composition of dust grains and their initial position in the ejecta.

In Table 3 we summarize the total mass $M_{\rm surv}$ of surviving dust 
in Type IIb SNRs at $t = 10^6$ yr for the CSM models considered in this 
paper.
For all CSM models except for the model A0.1, dust grains formed in the 
ejecta are completely destroyed before $t = 10^6$ yr.
Even for the model A0.1, the destruction efficiency $\eta$, defined as 
the ratio of $M_{\rm surv}$ to the initial dust mass $M_d$, is less 
than $10^{-3}$, in contrast to the case of SNe II-P with the thick 
hydrogen envelope, for which 2--80 \% of dust grains can survive for 
the ISM with $n_{\rm H,0} =$ 10--0.1 cm$^{-3}$ (Nozawa et al.\ 2007).
As shown above, the difference in the evolution of dust between SNe 
IIb and SNe II-P SNRs is caused by the thickness of outer envelope;
for SNe II-P retaining the massive hydrogen envelope, the radii of newly 
formed grains are larger than 0.01 $\mu$m, and the arrival time of the 
reverse shock at the dust-forming region inside the He core is retarded, 
both of which act against the destruction of dust grains by sputtering.
Therefore, how much amount of dust is injected into the ISM strongly 
depends on the type of SNe through the thickness of outer envelope, 
even if a large amount of dust grains can form in the ejecta of CCSNe.
From the results of the present calculations, we conclude that
low-mass envelope SNe such as SNe IIb and SNe Ib/c cannot be main 
contributors of interstellar dust as long as the gas density in the CSM 
is not too low ($n_{\rm H} \ll 0.1$ cm$^{-3}$).

\section{IR Thermal Emission from dust in Type IIb SNRs}

Dust grains bathed in the shocked gas inside SNRs are heated up by 
collisions with the hot gas and emit their thermal radiation at IR 
wavelengths.
If the collisional heating rate and radiative cooling rate of dust are 
balanced, the dust can acquire the equilibrium temperature.
However, in a rarefied hot plasma such as SNRs, small-sized dust grains 
undergo stochastic heating, which influences the emissivity and the 
resulting SED (e.g.,\ Dwek 1986).
In this section, based on the results of simulation of dust evolution 
presented in \S~3 and taking account into stochastic heating, we calculate 
the SED by thermal emission from shock-heated dust within SNRs.
The SNRs are placed at a distance of 3.4 kpc in order to compare with the 
observations of Cas A SNR in the next section.

\subsection{Calculation of stochastic heating}

The heating and thermal emission of dust are calculated by using the 
output of the dust evolution simulations:
density and temperature of the gas as well as size, velocity relative 
to the gas, and number density of each dust species in each spatial bin 
as a function of time.
We employ a Monte-Carlo method to calculate the temperature distribution 
functions of dust grains by stochastic heating.
In the calculations, we consider only collisions with electrons for the 
heating of dust;
under the present situation in which dust grains are quickly trapped in 
the hot gas, collisions with ions are much less frequent, and the 
heating of dust is dominated by collisions with electrons.

The fraction of incident energy that is deposited into a grain is 
calculated following the prescription by Dwek (1987a), and the stopping 
range of an electron for each grain species is calculated by the 
approximation formula which we derived based on those by Tabata et 
al.\ (1972) and Iskef et al.\ (1983).
The heat capacity of each grain species is taken from Takeuchi et al.\ 
(2005), and the optical constants are taken from 
the following sources: 
C (Edo 1983),
Al$_2$O$_3$ (Toon et al.\ 1976 for $\lambda \le 8$ $\mu$m and Begemann 
et al.\ 1997 for $\lambda > 8$ $\mu$m),
MgSiO$_3$ (Dorschner et al.\ 1995, but the optical constants of 
Mg$_2$SiO$_4$ are used for $\lambda \le 0.3$ $\mu$m, see Hirashita et al.\ 
2008), 
Mg$_2$SiO$_4$ (Semenov et al.\ 2003),
SiO$_2$ (Philipp 1985),
MgO (Roessler \& Huffman 1991),
FeS (Semenov et al.\ 2003),
Si (Piller 1985).
We re-bin the size of dust into 12 bins from 1 \AA~to 0.1 $\mu$m and 
divide the shocked region into 20 spatial bins weighted by the mass of 
the gas, because the calculation is too time-consuming.
We have verified that even if the number of re-binned size (spatial) 
bin increases to 20 (50), the resulting SEDs are virtually the same.

\subsection{Time evolution of thermal emission from dust}

Figures 11--14 show the time evolution of SEDs by thermal emission from 
shocked dust for different CSM models every 200 yr in the range from 200
yr to 1800 yr after the explosion.
In these figures, the solid curves display the SEDs taking into account 
stochastic heating, and the dashed curves the SEDs calculated by assuming 
the equilibrium without stochastic heating.
We can see that the SEDs taking account of the stochastic heating are 
strongly different from those with the equilibrium temperatures, 
especially at wavelengths shorter than $\lambda = 20$ $\mu$m.
This implies that the effect of stochastic heating is very important in 
modelling the IR SED by dust emission within SNRs, as demonstrated by 
Dwek (1986).

For the model A1 with the uniform gas density (Fig.\ 11), the reverse 
shock hits the He core at $t = 155$ yr after the explosion.
The mass of C grains formed in the outer C-rich region of the He layer 
is too small ($<$$5 \times 10^{-4}$ $M_\odot$) to emit the thermal 
emission significantly during $t=$ 200--400 yr.
At $t > 400$ yr, thermal emission from a large amount of C grains 
formed in the inner C-rich part of the He layer always dominates the SED, 
because the gas density in the SNR is not high enough to destroy C 
grains efficiently.
The reverse shock reaches the O-rich layer at $t \simeq 600$ yr, but 
the emission features of silicate and oxide grains are inconspicuous,
being overwhelmed by thermal emission from C grains.

For the higher CSM density of the model A10 (Fig.\ 12), C grains are 
more efficiently destroyed by sputtering than those in the model A1, 
and the emission features from other dust species get prominent with time. 
The reverse shock encounters the O-rich layer at $t = 270$ yr, and the 
emission feature of Mg-silicates (MgSiO$_3$ and Mg$_2$SiO$_4$) around 
$\lambda = 10$ $\mu$m begins to be noticeable at $t=600$ yr.
The feature of SiO$_2$ at $\lambda = 20$ $\mu$m is outstanding after 
$t=800$ yr, and the emission features from Si and FeS grains appear 
after $t=1200$ yr.
Because of the efficient destruction of dust, the thermal emission
continues to decrease with time.

In the case of the model B30 with the stellar wind's density profile 
(Fig.\ 13), the reverse shock collides with the He core at $t = 63$ yr, 
which is earlier than in the model A10.
Thermal emission from C grains dominates the SED by $t \sim 1000$ yr, 
and after then the contributions from Mg-silicates become recognizable.
However, stochastic heating of small C grains still contributes to the 
SED at $\lambda \la 10$ $\mu$m, because some C grains can survive even 
after $t = 1000$ yr owing to the low gas density ($n_{\rm H} = 1$ 
cm$^{-3}$) at a large distance in comparison with the model A10.
We can see that the flux densities significantly increase between 1200 
yr and 1400 yr in this model.
This is due to the fact that during this time interval, the reverse 
shock has swept up the ejecta region where a large amount of silicate 
grains are formed ($M_r \simeq 1.8$ $M_\odot$, see Fig.\ 3{\it a}).
The dust grains swept up by the shocks continue to be eroded by 
sputtering, while the reverse shock continues to intrude into the 
inner He core and to supply newly formed dust grains to the shocked gas.
Thus, the time evolution of SEDs is determined by the balance between 
the amount of dust destroyed in the hot gas and the amount of dust 
swept up by the reverse shock.

For the model B120 (Fig.\ 14), the reverse shock can quickly propagate 
into the ejecta, and arrives at the He core, O-rich layer, Si-rich 
layers at $t =$ 48 yr, 150 yr, and 560 yr, respectively.
Thus, we can see the quick evolution of the SED characterized by the 
features of various grains species.
Since the gas density is significantly high, especially at an early 
phase, C grains are quickly destroyed as can be seen from the evolution 
of SED at $\lambda \la 8$ $\mu$m, and the excess by stochastic heating 
of small C grains disappears at $t \ga 600$ yr.

The results of calculations clearly demonstrate that the time evolutions
of IR emission from dust heated up in SNRs are sensitive to the density 
of gas as well as the density profile in the surrounding medium.
The ambient gas density controls the passage of the reverse shock into 
the ejecta and affects the destruction by sputtering and stochastic 
heating of dust in the hot gas.
Hence, the resulting IR SED well reflects the evolution of dust in SNRs 
and also holds key information on the density structure in the ambient 
medium.
In other words, the temporal evolution of SED from dust in shock-heated 
gas can be a very useful probe for investigating the properties of dust 
in SNRs as well as the density structure in the CSM and thus the 
mass-loss history of the progenitor star during the late phase of its 
evolution.

\section{Applications to the Cas A SNR}

In this section, we apply the results of calculation for dust evolution
and the resulting SEDs presented in \S~3 and \S~4 to Cas A SNR, to give 
an insight into the composition and mass of dust residing in Cas A SNR 
and to get an implication on the mass-loss rate of the progenitor.
After briefly summarizing the observed characteristics of Cas A relevant 
to this study, we present the comparison of the calculated SEDs with the
observed one.

\subsection{Cas A SNR}

Cas A is one of the best studied Galactic SNRs at a broad range of 
wavelengths spanning from radio to $\gamma$-ray. 
Its estimated age is $\sim$330 years (Fesen et al.\ 2006a; Thorstensen 
et al.\ 2001), and the distance to Cas A is determined to be 3.4 kpc 
from the proper motion of fast moving knots (FMKs, Reed et al.\ 1995).
The X-ray observations with the {\it Chandra} have resolved the radii
of the reverse and forward shocks to be $95'' \pm 10''$ and 
$153'' \pm 12''$, respectively (Gotthelf et al.\ 2001), which correspond 
to $R_{\rm rev}^{330} = 1.57 \pm 0.17$ pc and 
$R_{\rm forw}^{330} = 2.52 \pm 0.19$ pc at a distance of 3.4 kpc.

Recently Krause et al.\ (2008) obtained an optical spectrum of Cas A SN 
around the maximum brightness via the scattered light echo, and concluded 
that the Cas A SN was of Type IIb, originated from a supergiant that had 
lost most of its hydrogen envelope prior to the explosion as has been 
suggested by the close binary scenario for SNe IIb (e.g., Nomoto et al.\ 
1995).
The numerical simulations (Chevalier \& Oishi 2003; Young et al.\ 2006) 
and the discovery of FMKs containing some hydrogen (Fesen \& Becker 1991;
Fesen 2001) also seem to strongly support this conclusion.

The presence of dust formed in the ejecta has been confirmed by the IR
observations with the {\it IRAS} (Dwek et al.\ 1987b), the {\it ISO} 
(Lagage et al.\ 1996), and the {\it Spitzer} (Hines et al.\ 2004).
The IR spectra of FMKs taken by the {\it ISO} allowed us to examine the 
composition of dust, leading to the conclusion that the main species of 
newly formed dust in Cas A are silicate and/or oxide grains 
(Arendt et al.\ 1999; Douvion et al.\ 2001).

There has been a continuing controversy about the amount of dust residing 
in Cas A SNR;
the dust mass derived from the observations described above was smaller 
than $3 \times 10^{-3}$ $M_\odot$. 
Rho et al.\ (2008) estimated the mass of dust to be 0.02--0.054 $M_\odot$
from the spectral mapping covering the entire region of Cas A by the 
{\it Spitzer}.
Based on the submillimeter (submm) observation with the {\it SCUBA},
Dunne et al.\ (2003) have reported the detection of 2--4 $M_\odot$ of 
cold dust in Cas A.
Dwek (2004) argued that this cold dust emission is ascribed to metallic
needles of $<$$10^{-3}$ $M_\odot$ formed in the ejecta.
On the other hand, the MIPS 160 $\mu$m map of Cas A, complemented by CO 
and OH line observations, have demonstrated that the detected submm 
emission is originated from interstellar dust in the foreground molecular 
clouds (Krause et al.\ 2004), and the mass of cold dust associated with 
Cas A is estimated at most to be 0.2 $M_\odot$ (Krause et al.\ 2004) and 
1.5 $M_\odot$ (Wilson \& Bartlra 2005).

Although the existence of a peculiar species of cold dust with $\sim$1 
$M_\odot$ is also proposed from submm polarization (Dunne et al.\ 2009), 
we do not here consider the measurements at wavelengths longer than 160 
$\mu$m.
In what follows, we compare the calculated SEDs with the 
synchrotron-subtracted IR flux densities at $\lambda =$ 8--100 $\mu$m 
taken from the photometric observations tabulated in Hines et al.\ 
(2004).

\subsection{Comparison with the observations of Cas A}

\subsubsection{IR emission from shock-heated newly formed dust}

We first compare the results for the uniform CSM with the observations 
of Cas A.
For $n_{\rm H,0} = 1$ cm$^{-3}$, IR emission is dominated by 
stochastically heated small C grains throughout the age of the remnant, 
as shown in \S~4.2.
This causes the excess emission over the observed SED at $\lambda \la 15$ 
$\mu$m until $t= 1000$ yr, and the calculated SED cannot reproduce the 
observational data.
The fact that the C grain-dominated IR SED does not match with the IR 
measurements of Cas A is also pointed out by Bianchi \& Schneider (2007).

For the case of $n_{\rm H,0}$ = 10 cm$^{-3}$, the calculated SED at 
$t = 800$ yr is in good agreement with the IR observations of Cas A 
(see Fig.\ 12).
The mass of shock-heated dust at $t=800$ yr is 0.048 $M_\odot$, which is 
consistent with the dust mass of 0.02--0.054 $M_\odot$ in Cas A estimated 
from the observations with the {\it Spitzer} (Rho et al.\ 2008).
The comparison strongly suggests that the dominant species of dust 
responsible for the observed SED are silicate grains and that the reverse 
shock has already reached the inner O-rich layer in the ejecta.
However, the remnant age of 800 yr is much older than the age of Cas A, 
and the SED at $t =330$ yr does not agree with the observations.
We also found that any uniform CSM models with gas density higher than 
$n_{\rm H,0} = 10$ cm$^{-3}$ cannot reproduce the observed SED, partly 
because the dust grains are quickly destroyed and partly because the 
temperature of dust is too high.
Thus, we conclude that thermal emission of dust grains heated up in SNRs 
expanding into the uniform CSM cannot explain the observed IR SED of 
Cas A.

Next we move to the comparison of the results for the CSM density 
structure of $\rho \propto r^{-2}$.
Figures 15{\it a}--15{\it c} present the calculated SEDs at $t = 330$ yr 
for $n_{\rm H,1}$ = 30, 120, and 200 cm$^{-3}$.
In these figures, we also show the contribution of each dust species 
to the total flux densities.
Note that the SEDs calculated without stochastic heating are not shown 
in these figures, since they can never reproduce the observations of 
Cas A for any CSM models.

The case of $n_{\rm H,1} = 30 $ cm$^{-3}$ corresponds to 
$\dot{M} = 2 \times 10^{-5}$ $M_\odot$ yr$^{-1}$ for $v_w = 10$
km s$^{-1}$, and the density structure has been suggested by Chevalier 
\& Oishi (2003) to reproduce the observed radii of the reverse and forward
shocks derived from the X-ray observations (Gotthelf et al.\ 2001).
Our simulation also approximately reproduces the observed shock radii for 
Cas A: 1.5 pc for the reverse shock and 2.2 pc for the forward shock.
However, in this case, the reverse shock has not yet reached the O-rich 
layer at $t = 330$ yr (see Table 3).
The calculated flux densities from the shock-heated C grains of
$\sim$0.05 $M_\odot$ are much higher than the observations at the whole 
mid-IR wavelengths.

With $n_{\rm H,1} = 120$ cm$^{-3}$, the calculated SED reasonably 
reproduces the observed SED at $\lambda \la 30$ $\mu$m, although there is 
some discrepancy at $\lambda \simeq 12$ $\mu$m.
The SED also predicts the considerably lower flux densities at 
$\lambda \ga 50$ $\mu$m, which can be recovered by considering the 
unshocked dust as will be discussed later.
The dust species most contributing to the SED is MgSiO$_3$, which is one 
of the major grain species in Cas A presumed from the IR spectra 
(Douvion et al.\ 2001; Rho et al.\ 2008).
Mg$_2$SiO$_4$, SiO$_2$, and MgO grains more or less contribute to the SED, 
and C grains partly contribute at $\lambda \la 15$ $\mu$m as well.
The total mass of shocked dust grains is 0.008 $M_\odot$, which is smaller 
than the estimate by Rho et al.\ (2008) but is larger than those by the 
other IR observations.

For $n_{\rm H,1} = 200$ cm$^{-3}$, the calculated SED does not match 
with the observations, especially at $\lambda =$ 20--60 $\mu$m. 
In this case, the reverse shock invades deeper layers in the ejecta and 
sweeps up larger mass (0.028 $M_\odot$) of silicate grains, compared 
with the case of $n_{\rm H,1} = 120$ cm$^{-3}$.
Therefore, the derived SED produces the much more flux densities than 
the observations.
We also calculate the thermal emission from shocked dust for the denser 
CSM than $n_{\rm H,1} = 200$ cm$^{-3}$, but the outcomes suffer from 
much less flux densities at $\lambda \ga 30$ $\mu$m because of the 
efficient destruction of dust and the relatively high temperature of 
dust.
In addition, in such dense cases, the shock radii are much smaller than
the observations.
Hence, we conceive that the cases of $n_{\rm H,1} \ga 200$ cm$^{-3}$ 
are unlikely as the density structure around Cas A.

In summary, our results indicate that the stellar wind's density profile
with $n_{\rm H,1} \simeq 120$ cm$^{-3}$ is most favorable to account for 
the observed IR emission at the age of Cas A.
The corresponding mass-loss rate $8 \times 10^{-5}$ $M_\odot$ yr$^{-1}$
is at most a factor 4 higher than the mass-loss rates estimated for SNe 
IIb: (3--4)$\times 10^{-5}$ $M_\odot$ yr$^{-1}$ (Suzuki \& Nomoto 1995;
Fransson et al.\ 1996) deduced for SN 1993J 
(at $r \simeq 1 \times 10^{15}$ cm) and $2 \times 10^{-5}$ $M_\odot$ 
yr$^{-1}$ for Cas A (Chevalier \& Oishi 2003). 
In view of the uncertainties involved in the density gradient and 
clumpiness in the CSM, these estimated mass loss rates can be regarded
as being consistent.
On the other hand, the result for $n_{\rm H,1} = 120$ cm$^{-3}$ predicts 
quite smaller shock radii than those for Cas A (see Table 3).
The discrepancy in the shock radii as well as the mass-loss rate could be 
considered to be partly caused by the difference in the explosion energy 
and the thickness of the hydrogen envelope of the SN model used in this 
study.

It should be emphasized that the IR emission from newly formed dust grains
that are heated up in the hot plasma inside SNRs can reproduce the mid-IR 
($\lambda =$ 8--30 $\mu$m) SED of Cas A without any fine-tuning.
On the other hand, the obtained results for $n_{\rm H,1} = 120$ cm$^{-3}$ 
significantly underestimate the flux densities at longer wavelengths 
($\lambda \ga 50$ $\mu$m).
Although the thermal emission from cooler dust in the dense optically 
emitting knots is expected to mitigate this difference, modelling of such
a dust emission may need a sophisticated radiative transfer simulation and 
is beyond the scope of this paper. 
Hereafter, we discuss the contributions from the following two potential 
sources to the observed SED at $\lambda \ga 50$ $\mu$m:
(1) the circumstellar (CS) dust grains that was formed in the mass-loss 
wind during the evolution of the progenitor star and then were swept up 
by the forward shock, and 
(2) the newly formed dust that were formed in the ejecta but have not 
been yet swept up by the reverse shock.

\subsubsection{Contribution from preexisting CS dust}

In order to examine the contribution from the preexisting CS dust, we 
simulate the destruction and heating of the shocked CS dust for the 
density structure of the CSM corresponding to the mass-loss rate of 
$8 \times 10^{-5}$ $M_\odot$ yr$^{-1}$.
The composition of the CS dust is considered to be either carbonaceous 
or silicate depending on the C/O ratio in the stellar wind.
Carbonaceous grains condense in C/O $> 1$, which is realized in the winds 
of carbon-rich Wolf-Rayet stars. 
Silicate grains are expected to form in the wind of a supergiant, where 
the abundance ratio of heavy elements is nearly solar with C/O $< 1$.
Although a supergiant is a more likely progenitor of Cas A (\S~5.1), we 
consider here, as the dust composition in the CSM, both amorphous carbon 
and astronomical silicate, for which the optical constants are taken from 
Edo (1983) and Draine \& Lee (1984), respectively.
For the size distribution of dust, we adopt the MRN size distribution 
($n(a)da \propto a^{-3.5}da$, Mathis et al.\ 1977) with the upper and 
lower limits of $a=0.001$ $\mu$m and $0.5$ $\mu$m, respectively.
The dust-to-gas mass ratio in the CSM is assumed to be constant and is
taken as a free parameter.

Figure 16 illustrates the contribution from CS carbon 
(Fig.\ 16{\it a}) and astronomical silicates (Fig.\ 16{\it b}) 
for the case of $n_{\rm H,1} = 120$ cm$^{-3}$, where the ejecta-dust and 
CS dust components are depicted by the dotted and dashed lines, 
respectively.
The dust-to-gas mass ratios adopted here are $1.6 \times 10^{-4}$ for 
carbon and $2.8 \times 10^{-4}$ for silicate because higher dust-to-gas 
mass ratios cause the excess emission over the observed SED 
at $\lambda \la 50$ $\mu$m.
At $t = 330$ yr, the masses of shock-heated CS carbon and silicate 
grains are $1.4 \times 10^{-3}$ $M_\odot$ and $1.8 \times 10^{-3}$ 
$M_\odot$, respectively, and both grain species have typical 
temperature of $\simeq$100 K.
Thus, we can see that both grain species can only contribute to the SED 
at $\lambda \la 50$ $\mu$m with the peaks around $\lambda = 30$ $\mu$m, 
and cannot compensate the flux densities at $\lambda \ge 50$ $\mu$m.
This means that the observed SED cannot be improved by the contributions 
from CS grains even if we consider any mixture of these CS grains and 
dust-to-gas mass ratios different from those considered here. 
Therefore, we suppose that the IR emissions from shocked CS grains 
should not significantly contribute to the IR SED of Cas A
and that a small content of CS dust may be due to low condensation 
efficiency of dust in the mass-loss wind and/or due to partial 
destruction of CS dust by the strong radiation associated with the shock
breakout at the explosion (e.g., Blinnikov et al.\ 1998).
We emphasize that the thermal emission from CS dust alone can never 
reproduce the observations of Cas A for any models considered in this 
paper.

\subsubsection{Contribution from unshocked ejecta-dust}

As shown in \S~5.2.1, in the case of $n_{\rm H,1} = 120$ cm$^{-3}$, only 
0.008 $M_\odot$ of newly formed dust grains is heated in the shocked gas 
at $t = 330$ yr, and 0.072 $M_\odot$ of dust grains remains cold without 
being swept up by the reverse shock (see Table 3).
These unshocked dust grains are expected to be heated up by absorbing 
the X-ray and UV/optical photons generated from the shock-heated plasma 
and dense knots, and they may radiate their thermal emission with the 
higher temperature than that ($\sim$20 K) in the diffuse ISM and 
contribute to the SED at IR wavelengths.
In fact, the presence of cold dust in the unshocked ejecta of Cas A SNR 
has been suggested from the association of a ``featureless dust'' 
component with the Si II emission from the central part of the remnant 
(Rho et al.\ 2008).
Since the dedicated calculations of radiative transport at X-ray to 
optical wavelengths is beyond the scope of this paper, we evaluate the 
contribution of the unshocked dust to the IR SED, assuming the same 
temperature for all unshocked grains.

Figure 17 shows the contribution from unshocked newly formed dust,
together with the contribution of shock-heated warm dust at 330 yr 
(Fig.\ 17{\it a}) and 340 yr (Fig.\ 17{\it b}) for $n_{\rm H,1} = 120$ 
cm$^{-3}$.
The age of 340 yr is the expected age of Cas A in the case where 
the forward shock has not been decelerated (Thorstensen et al.\ 2001; 
Fesen et al.\ 2006a).
As can be seen from the figures, if the unshocked dust grains are 
heated up to a temperature around 40 K, their thermal emission can 
significantly contribute to flux densities at $\lambda \ga 50$ $\mu$m 
and can successfully reproduce the observations of Cas A.
Thus, the IR SED observed for Cas A can be naturally explained by the 
shocked warm component and unshocked cold component of newly formed dust 
obtained by a series of our calculations.
We also mention that the calculated IR emission from shocked warm 
dust at $t = 340$ yr slightly increases from that at $t = 330$ yr, 
giving a better fit at $\lambda \simeq 20$ $\mu$m.
This indicates that the IR flux from Cas A can change during a short 
time interval of $\sim$10 yr.

Finally, it should be pointed out here the followings:
We show that the IR observations of Cas A are successfully reproduced 
by 0.008 $M_\odot$ shocked dust and 0.072 $M_\odot$ unshocked dust, 
considering the destruction and heating of newly formed dust in the SNR 
for $n_{\rm H, 1} = 120$ cm$^{-3}$.
Although the derived dust mass in Cas A SNR is superficially large enough 
to explain a vast amount of dust observed in the host galaxies of 
high-redshift quasars at $z > 5$, these dust grains would be completely 
destroyed in $2 \times 10^4$ yr and could not be supplied into the ISM, 
as demonstrated in \S~3.2, if the CSM would be more or less spherical.
We also should note that if the CSM would not be spherical, a part of
dust grains in the unshocked ejecta could be ejected into the ISM.

\section{Conclusion}

We investigate the formation of dust in the ejecta of a hydrogen-poor 
SN IIb, adopting the SN model with $E_{\rm kin}=10^{51}$ ergs,
$M_{\rm eje} = 2.94$ $M_\odot$, and $M_{\rm env}= 0.08$ $M_\odot$.
Then we explore the evolution of newly formed dust in the hot gas inside 
the SNR by considering the two sets of density profile for the spherical 
CSM.
Based on the results of dust evolution in SNRs, we also calculate the 
time evolution of thermal emission from the shocked dust within the SNR 
and compare the results with the observations of Cas A.
The results of our calculations are summarized as follows.

(1) 
In the unmixed ejecta of the SN IIb, various dust species can condense 
depending on the elemental composition in each layer at 300--700 days 
after the explosion. 
The total mass of newly formed dust is as large as 0.167 $M_\odot$, and 
the composition and mass of dust formed in the SN IIb are similar to 
those in SNe II-P.
However, the mass of the hydrogen envelope of the SN IIb is too small
to significantly decelerate the expansion of the He core.
Thus, the gas density in the He core is too low for large-sized grains 
to be produced, and the resulting average radius of dust is confined to 
be smaller than 0.01 $\mu$m.
We propose that the size of dust formed in the SNe with the low-mass 
envelope such as SNe IIb as well as the envelope-stripped SNe Ib/Ic is 
much smaller than that formed in SNe II-P with the massive hydrogen 
envelope.

(2) 
The evolution of dust within SNRs is strongly affected by the mass of 
the outer envelope and the gas density in the CSM.
The less massive outer envelope and/or the denser CSM lead to the earlier 
arrival of the reverse shock at the He core, and causes the quicker and 
more efficient destruction of newly formed dust in the denser hot gas.
In the Type IIb SNR, as a result of the small size of dust and the early 
propagation of the reverse shock into the dust-forming region, all newly 
formed dust grains are destroyed in the shocked gas without being 
injected into the ISM for the ambient gas density of $n_{\rm H} > 0.1$ 
cm$^{-3}$.
If the CSM is more or less spherical, our results imply that low-mass 
envelope SNe do not play a major role in the dust enrichment in the ISM.

(3)
The time evolution of IR SEDs by thermal emission from shocked dust in 
SNRs well reflects the evolution of dust through the erosion by sputtering 
and stochastic heating.
The destruction and heating processes of dust in the hot gas are heavily 
affected by the ambient gas density that affects the passage of the
reverse shock into the ejecta.
Thus, the comparison of theoretically modelled SEDs and observed IR SEDs 
can be used to diagnose the size and mass of dust in SNRs and the density 
structure surrounding the SNe.
These in turn provide us with some hints on the formation process of dust 
in the ejecta and the mass-loss activity of the progenitor stars during 
the late evolution.

(4) 
The IR SED observed for Cas A can be reasonably reproduced by the 
shocked warm dust with $M_d = 0.008$ $M_\odot$ and unshocked cold 
dust with $M_d = 0.072$ $M_\odot$, both of which were newly formed in 
the ejecta.
The CSM density required to reproduce the observed SED corresponds to 
the steady stellar wind of a supergiant with the mass-loss rate 
of $8 \times 10^{-5}$ $M_{\odot}$ yr$^{-1}$.
This mass-loss rate is consistent with those estimated for the progenitors 
of SN 1993J and Cas A within a factor of 2--4.
It should be emphasized that these conclusions are the natural outcomes
derived by the self-consistent treatments of formation, destruction, and 
heating of dust given in this study.

The result of the dust formation calculation shows that the total mass 
of dust formed in the ejecta is 0.167 $M_\odot$, which is much larger than 
the mass estimate of dust deduced from the observations of nearby 
dust-forming SNe.
From the comparison of calculated SEDs with the observed SED, we predict
that in Cas A SNR about a half of newly formed dust grains has already
been swept up by the reverse shock and most of them had been destroyed, 
while the other half has not yet been swept up by the reverse shock.
The mass of this unshocked dust (0.072 $M_\odot$) is in good agreement 
with the upper mass limit of cold dust (0.2 $M_\odot$) in Cas A estimated 
by Krause et al.\ (2004).
On the other hand, our calculation demonstrates that all dust grains 
formed in Cas A might be completely destroyed later on and might not be 
injected into the ISM, if the CSM would be more or less spherical
(see below).
Thus, the SN IIb that formed Cas A might not be a representative of SNe 
as major sources of interstellar dust, but Cas A could have remained as a 
unique and ideal object for investigating formation and evolution 
processes of dust in SNe.

We should bear in mind that all calculations here are performed for the
spherically symmetric forward and reverse shocks propagating into the 
homogeneous CSM and stratified ejecta, respectively.
In the binary scenario for SNe IIb (Nomoto et al.\ 1995, 1996), the 
formation of aspherical CSM is predicted to be as follows.
The close binary consists of a massive SN IIb progenitor and a small-mass
companion star.
In the common envelope phase, the companion star spirals into the envelope
of the SN IIb progenitor, which causes extensive mass loss from the H-rich 
envelope of the progenitor in the direction of equatorial plane.
The resultant CSM might have a ring-like structure as seen in SN 1987A, 
a torus, or a thick disk.
In reality, the CSM around SN 1993J is considered to be more like torus 
shape (e.g., Fransson et al.\ 2005).
Then since the ejecta of polar direction cannot strongly interact with the 
CSM (e.g., Maeda et al.\ 2002), the ejecta of SN 1993J can be composed of 
the reverse-shocked ejecta and the unshocked ejecta.
Thus, if CSM is spherical, dust can be totally destroyed.
If CSM is non-spherical like SN 1993J, unshocked dust can partially be 
ejected.

In addition, the IR and X-ray observations of Cas A have shown that the 
ejecta of Cas A are very clumpy and asymmetric (Willingale et al.\ 2002;
Ennis et al.\ 2006; Smith et al.\ 2009) with the inverted compositions 
as represented by Fe-rich materials overtaking the outer layer (e.g., 
Hughes et al.\ 2000).
This indicates that some dust species may encounter the reverse shock at 
different times and different gas densities than suggested in this paper. 
Furthermore, if newly formed grains would reside in dense clumps, it 
might be possible that a part of them can survive the destruction and be 
injected into the ISM.
Although the mass of shocked and visible material is believed to roughly 
account for all the mass ejected in the explosion, the IR spectral 
mapping observations of the entire region of Cas A SNR suggest the 
presence of unshocked ejecta materials in the central region of the 
remnant (Rho et al.\ 2008; Smith et al.\ 2009).
However, it has been difficult to estimate how much amount of dust is 
there in the unshocked region from the observations covering the 
wavelengths of $\lambda \le 70$ $\mu$m, as pointed out by Rho et al.\ 
(2008).
Since the mass of dust in Cas A is likely to be dominated by the 
unshocked low-temperature dust, the comprehensive far-IR observations at 
$\lambda \ga 50$ $\mu$m are necessary to reveal the mass of dust formed 
in the Cas A SN.
\footnote{After this paper was submitted, an article by Sibthorpe et 
al.\ (2009) appeared on astro-ph (arXiv/0910.1094), reporting the 
presence of a cool emission component peaked near the center of Cas A 
in the 90 $\mu$m image obtained by the {\it AKARI}.
They suggest that this emission comes from unshocked newly formed dust
in Cas A and estimate its mass to be 0.03--0.06 $M_\odot$.}
Unfortunately, Cas A suffers from the contamination from the interstellar 
dust in the foreground molecular clouds, but the {\it Herschel} will be 
able to give more critical clues on the mass of cold dust in Cas A.

\acknowledgments

The authors are grateful to the anonymous referee for critical comments
that are useful for improving the manuscript.
T. N. and T. K. would like to acknowledge Th. Henning and all members for 
the kind hospitality during their stay at MPIA.
This research has been supported in part by World Premier International 
Research Center Initiative (WPI Initiative), MEXT, Japan, and by the 
Grant-in-Aid for Scientific Research of the Japan Society for the 
Promotion of Science (18104003, 19740094, 20340038, 20540226, 20540227,
20840007, 21840055) and MEXT (19047004, 20040004).


\clearpage

\begin{figure}
\epsscale{0.6}
\plotone{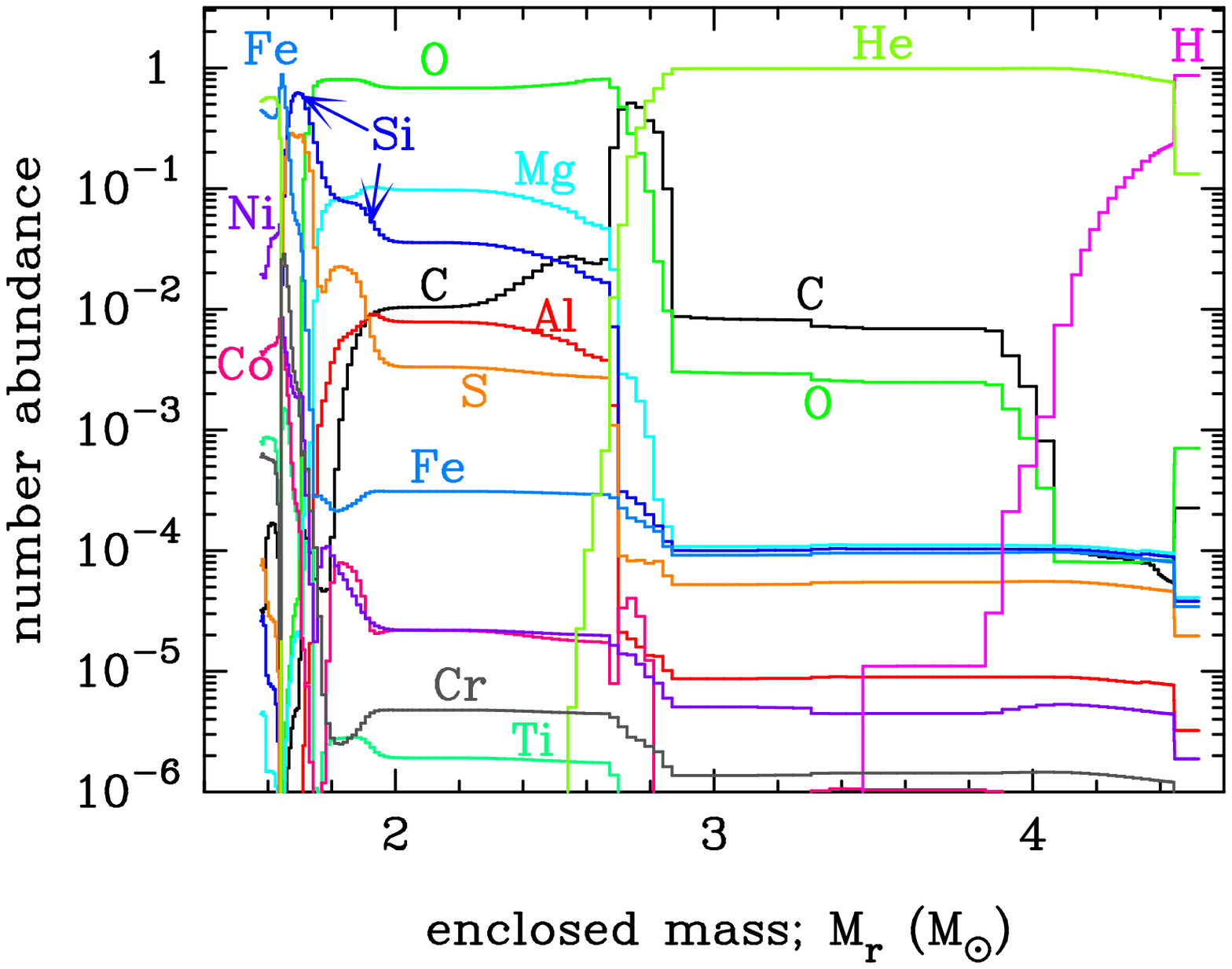}
\caption{
 Relative number abundance of elements in the ejecta of the SN IIb model 
 at 600 days after the explosion as a function of enclosed mass.
 The decay of radioactive elements is taken into account, and the original 
 onion-like structure is assumed to be preserved without any mixing of 
 elements.
 The progenitor mass of the SN IIb is 18 $M_\odot$, and the mass at the 
 explosion is 4.52 $M_\odot$.
 The mass cut is 1.58 $M_\odot$, which results in the ejecta mass of 2.94 
 $M_\odot$.
 The mass of the hydrogen envelope is 0.08 $M_\odot$.
\label{fig1}}
\end{figure}

\clearpage

\begin{figure}
\epsscale{0.6}
\plotone{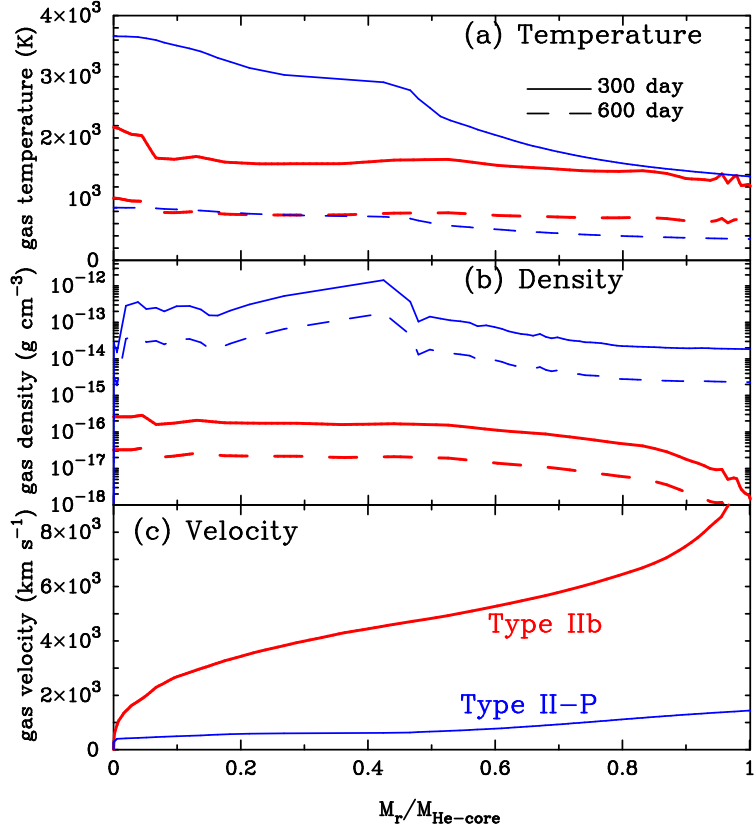}
\caption{
 Structures of ({\it a}) temperature and ({\it b}) density of the gas 
 within the He core at day 300 ({\it solid lines}) and day 600 ({\it 
 dashed lines}) after the explosion, and ({\it c}) the velocity 
 distribution for the SN IIb model ({\it thick lines}).
 For comparison, shown are those for the SN II-P model ({\it thin lines}) 
 with $M_{\rm eje} = 17$ $M_\odot$ and $M_{\rm env} = 13.2$ $M_\odot$
 (Umeda \& Nomoto 2002).
 The mass coordinate is normalized by the mass of the He core.
 Note that both SN models have $E_{\rm kin} = 10^{51}$ ergs and 
 $M(^{56}{\rm Ni}) = 0.07$ $M_\odot$.
\label{fig2}}
\end{figure}

\clearpage

\begin{figure}
\epsscale{0.5}
\plotone{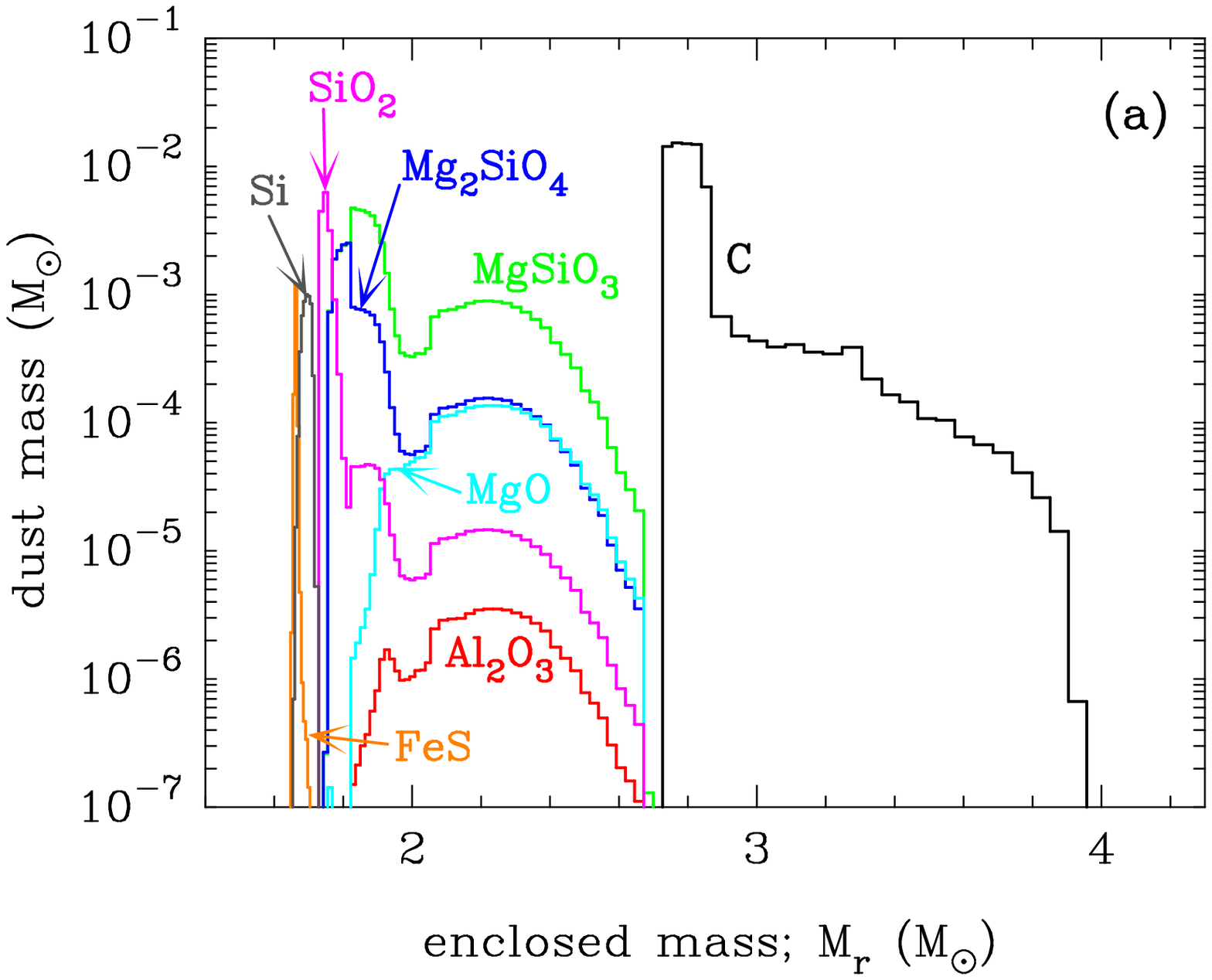}
\plotone{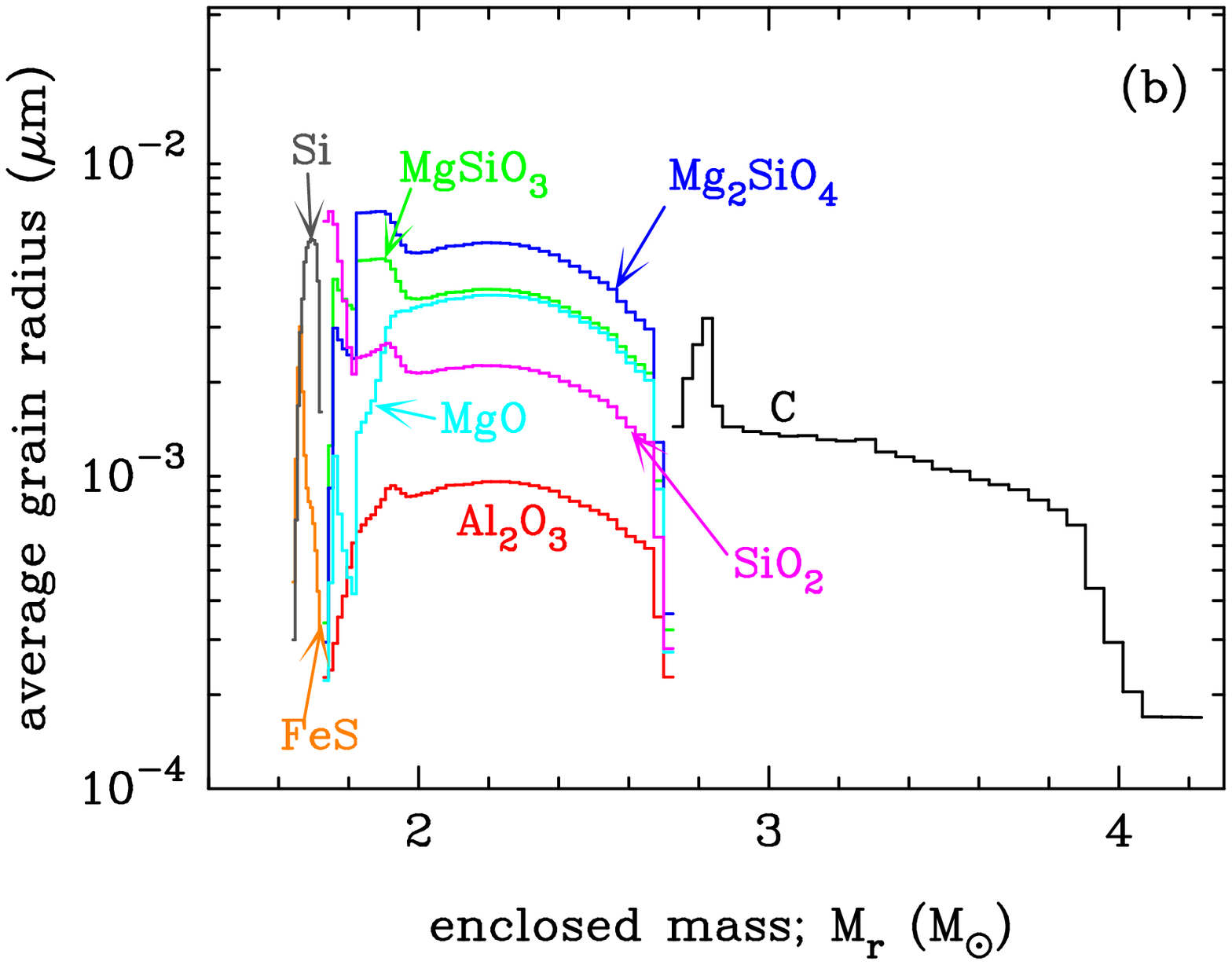}
\caption{
 ({\it a}) Mass distribution and ({\it b}) average radius of each dust 
 species formed in the ejecta of the SN IIb model as a function of 
 enclosed mass.
\label{fig3}}
\end{figure}

\clearpage

\begin{figure}
\epsscale{0.6}
\plotone{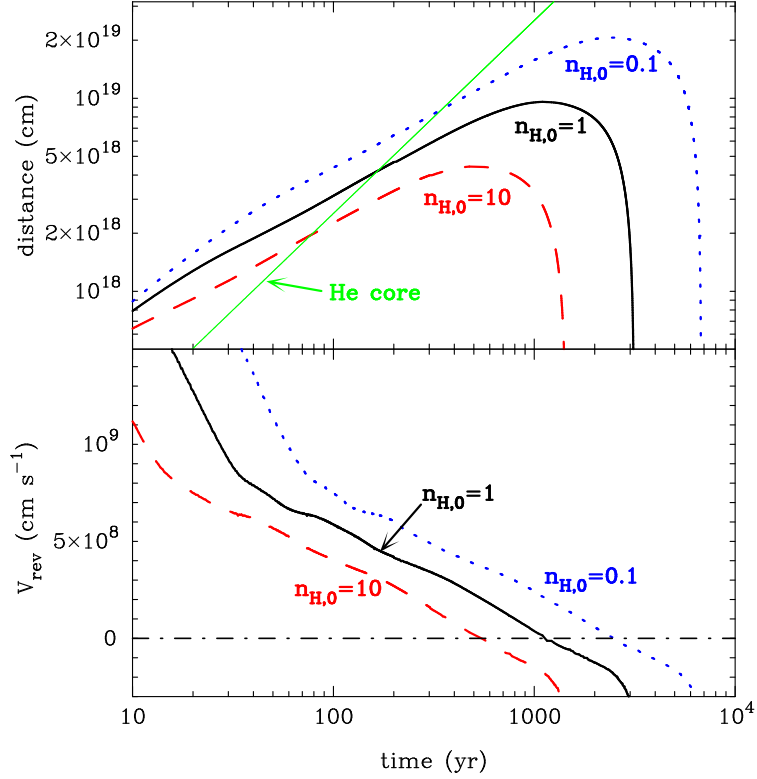}
\caption{
 Time evolution of radius ({\it upper panel}) and velocity ({\it lower 
 panel}) of the reverse shock for the uniform CSM models A0.1 
 ({\it dotted}), A1 ({\it solid}), and A10 ({\it dashed}) with 
 $n_{\rm H,0}=0.1$, 1, and 10 cm$^{-3}$, respectively. 
 The trajectory of the boundary between the He core and the hydrogen 
 envelope is depicted by the thin solid curve in the upper panel, 
 where the boundary is assumed not to be decelerated after the collision
 with the reverse shock for illustration. 
\label{fig4}}
\end{figure}

\clearpage

\begin{figure}
\epsscale{0.6}
\plotone{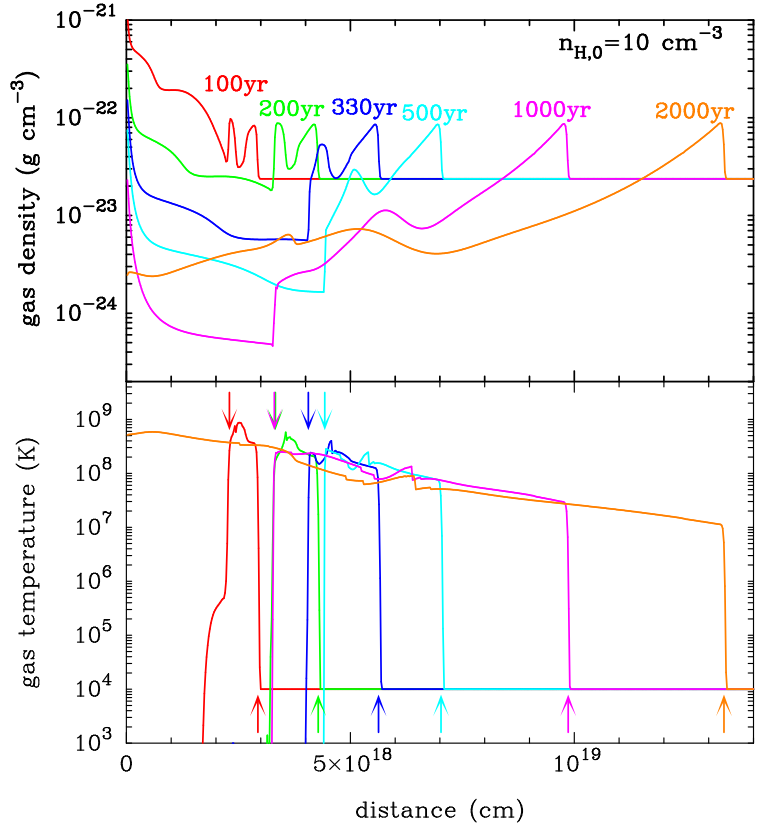}
\caption{
 Structures of density ({\it upper panel}) and temperature ({\it lower 
 panel}) of the gas at given times within the Type IIb SNR for the model
 A10 with $n_{\rm H,0}=10$ cm$^{-3}$. 
 The positions of the forward and reverse shocks are indicated by the 
 upward- and downward-pointing arrows in the lower panel, respectively.
\label{fig4}}
\end{figure}

\clearpage

\begin{figure}
\epsscale{0.6}
\plotone{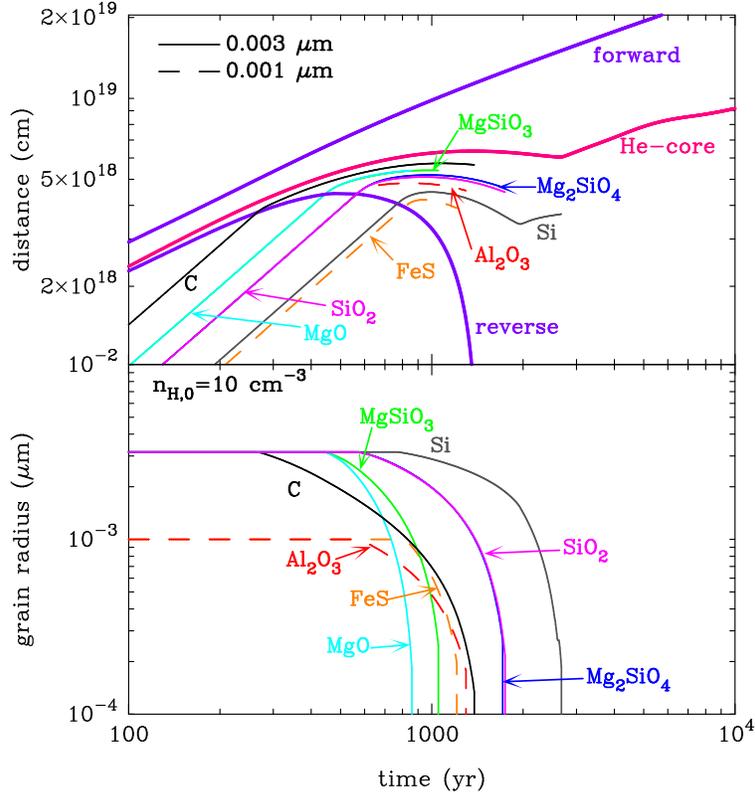}
\caption{ 
 Time evolutions of the positions ({\it upper panel}) and radii ({\it 
 lower panel}) of dust grains within the Type II SNR for the model A10.
 The initial radii of grains shown in the figures are 0.001 $\mu$m for 
 Al$_2$O$_3$ and FeS ({\it dashed lines)} and 0.003 $\mu$m for C, 
 MgSiO$_3$, Mg$_2$SiO$_4$, SiO$_2$, MgO, and Si ({\it solid lines}). 
 The trajectories of the reverse and forward shocks as well as the 
 position of the boundary between the He core and the hydrogen envelope 
 are depicted by the thick curves in the upper panel.
\label{fig5}}
\end{figure}

\clearpage

\begin{figure}
\epsscale{0.5}
\plotone{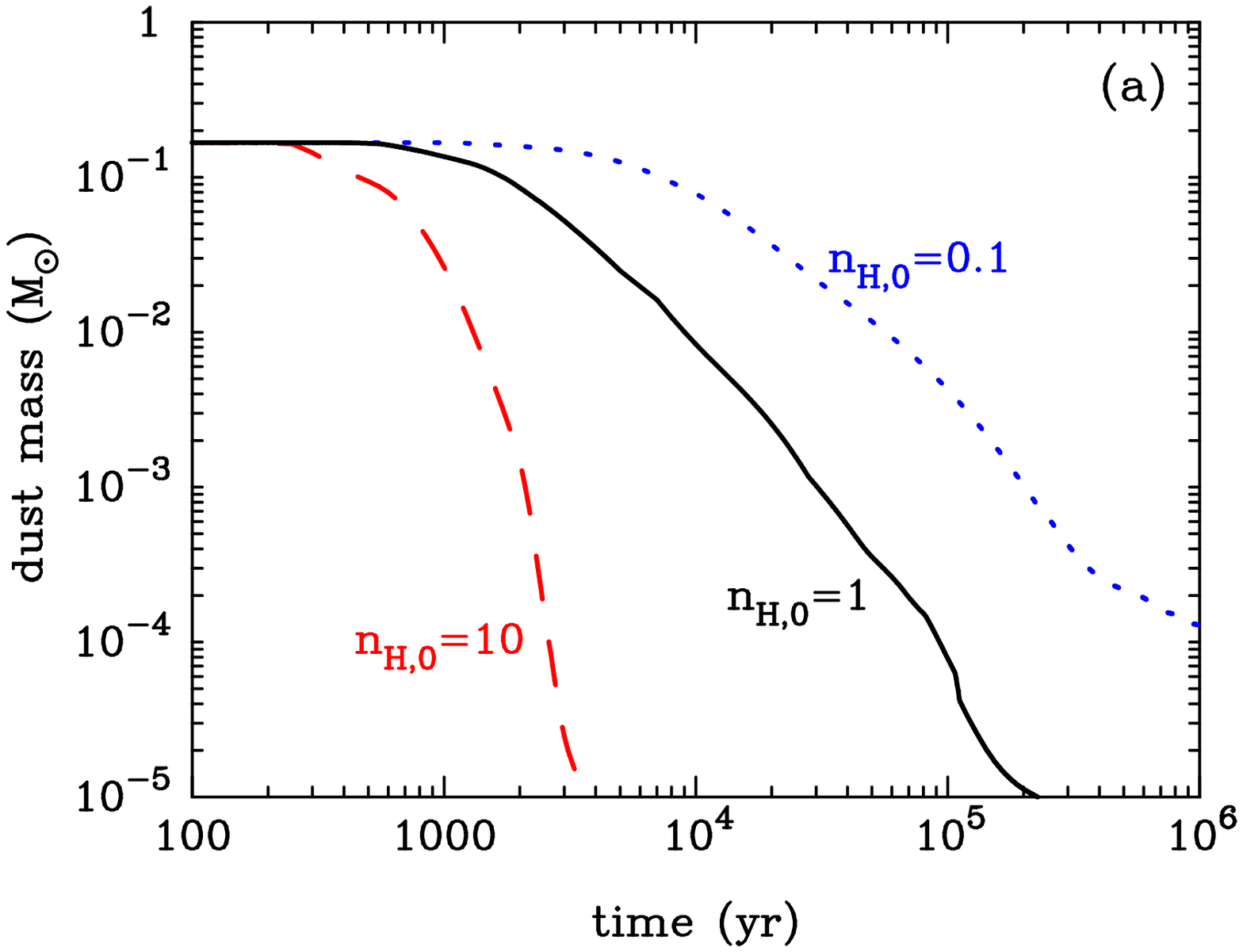}
\plotone{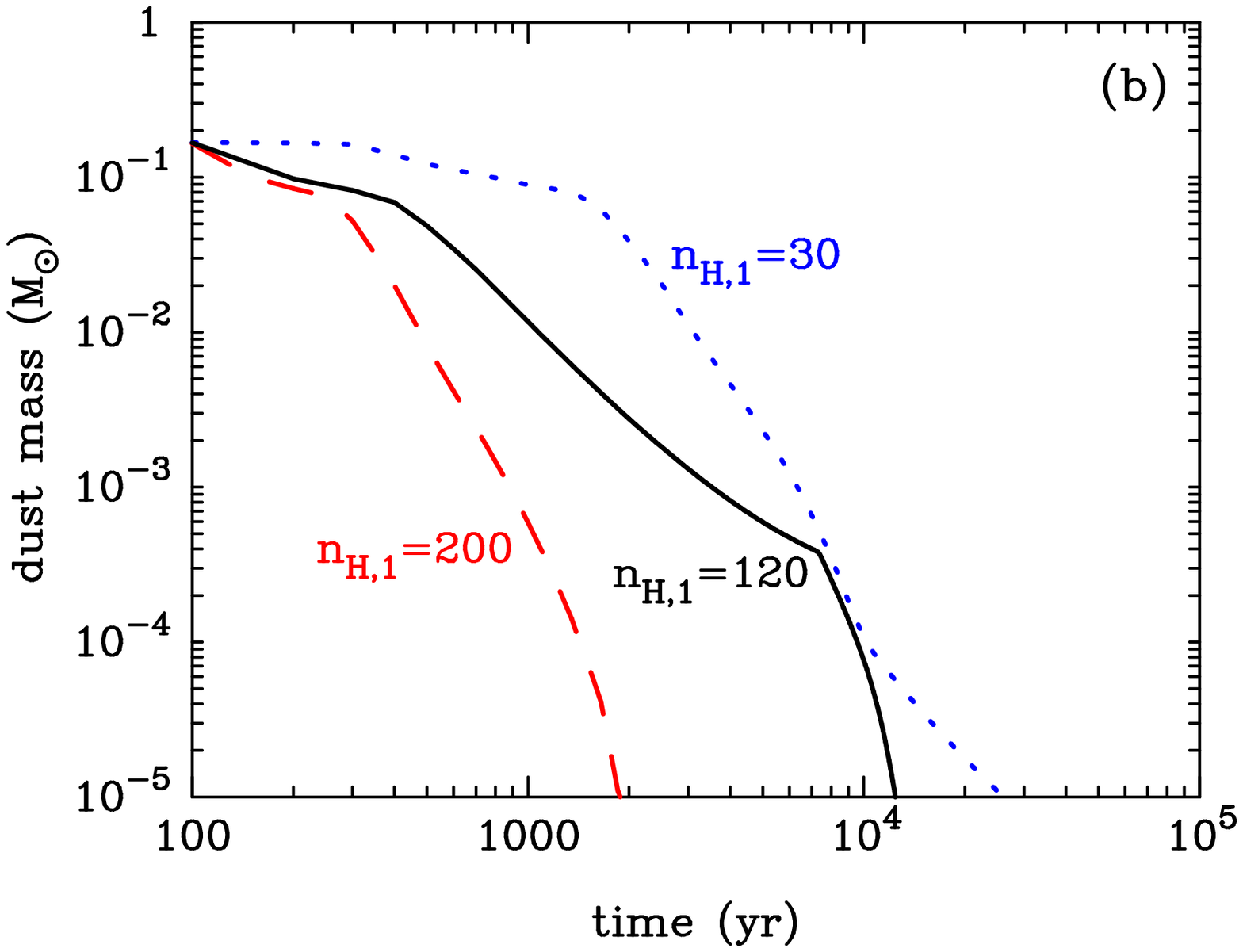}
\caption{ 
 Time evolutions of the total dust mass inside the Type IIb SNRs 
 running into ({\it a}) the uniform CSM with $n_{\rm H,0} = 0.1$ cm$^{-3}$ 
 ({\it dotted}), 1 cm$^{-3}$ ({\it solid}), and 
 10 cm$^{-3}$ ({\it dashed}), and ({\it b}) the stellar wind CSM 
 with $n_{\rm H,1} = 30$ cm$^{-3}$ ({\it dotted}), 
 120 cm$^{-3}$ ({\it solid}), and 
 200 cm$^{-3}$ ({\it dashed}).
\label{fig6}}
\end{figure}

\clearpage

\begin{figure}
\epsscale{0.5}
\plotone{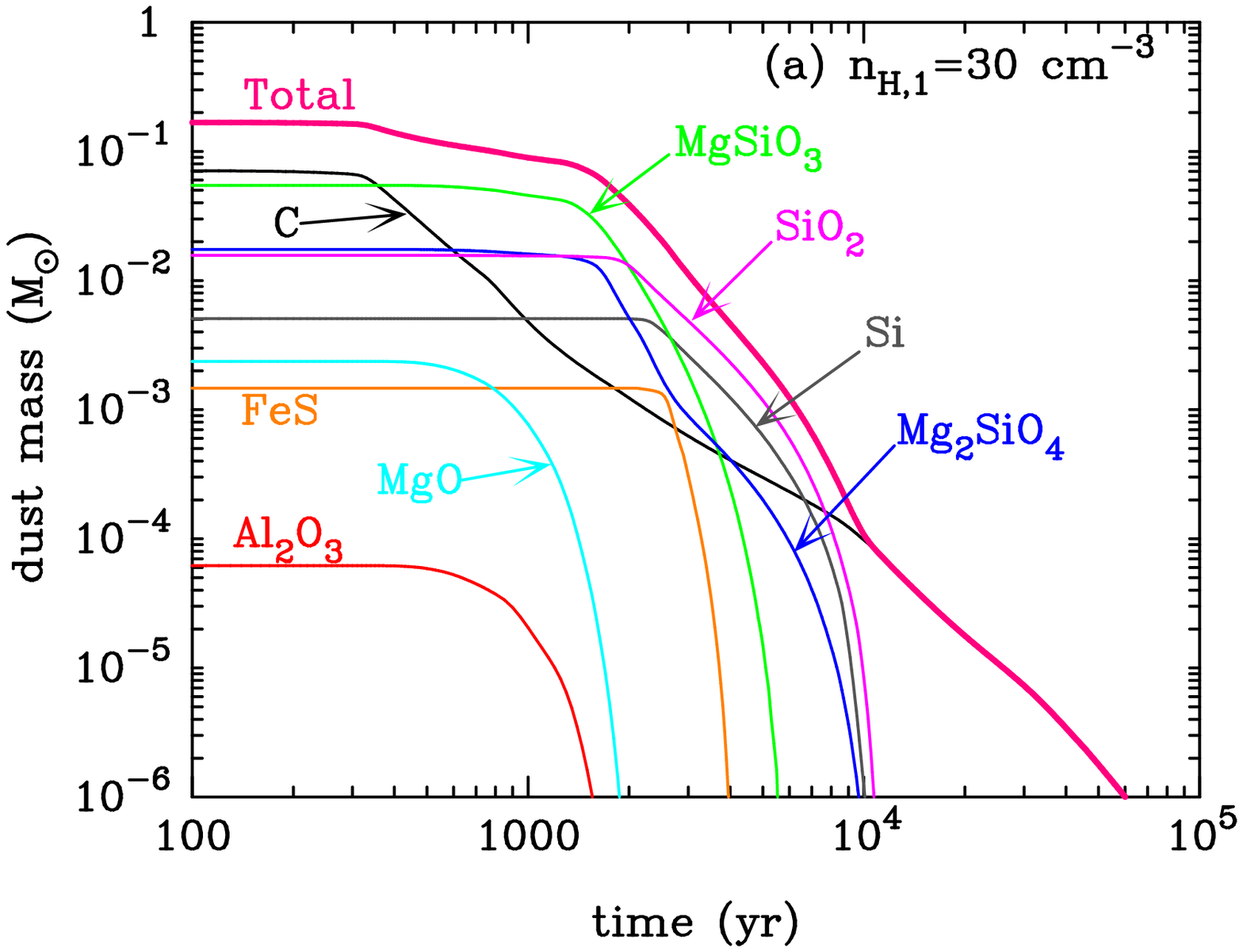}
\plotone{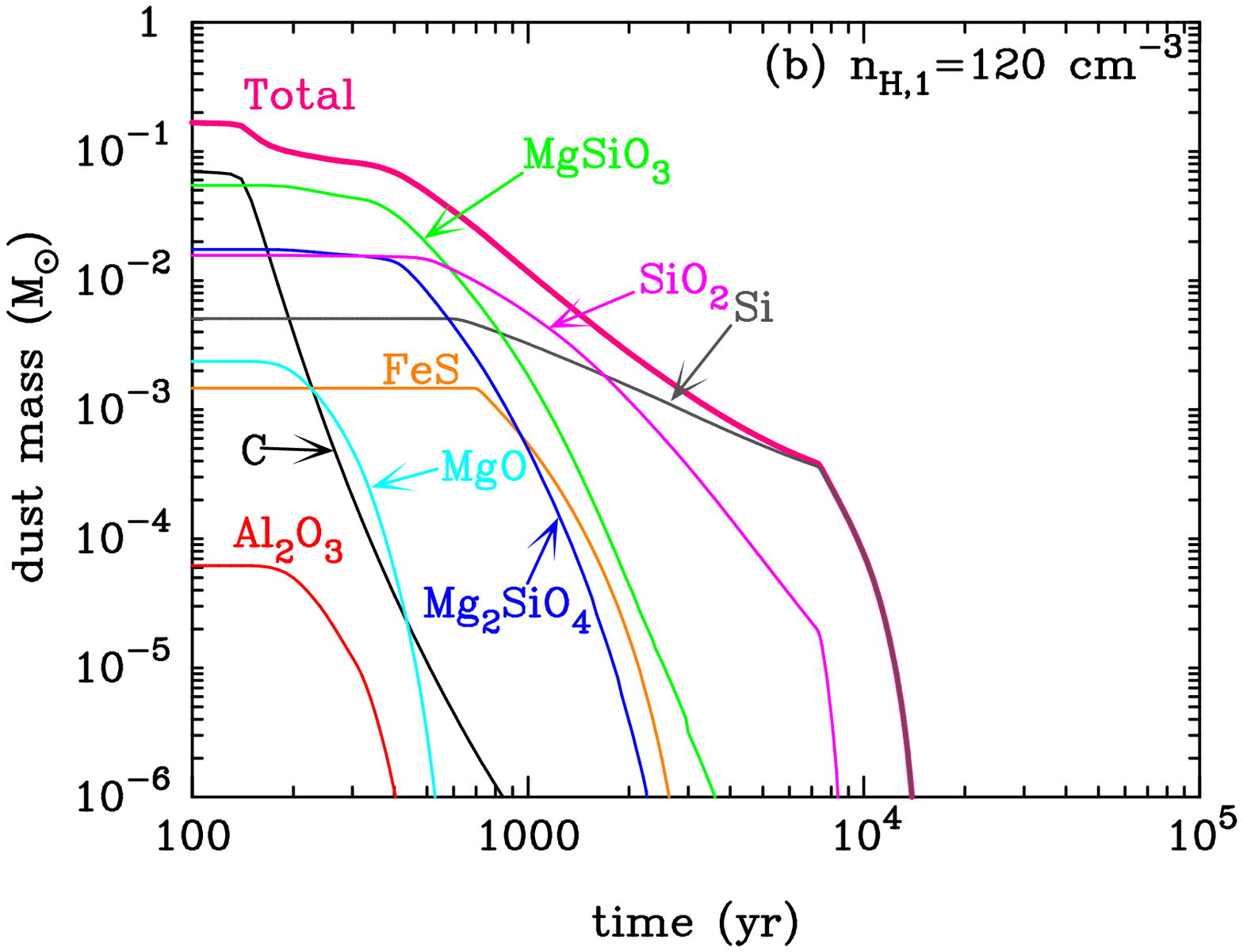}
\caption{ 
 Time evolution of the mass of each dust species inside the Type IIb 
 SNRs running into the stellar wind CSM with the density profile of 
 $\rho \propto r^{-2}$ for ({\it a}) $n_{\rm H,1} = 30$ cm$^{-3}$ and 
 ({\it b}) $n_{\rm H,1} = 120$ cm$^{-3}$.
 The thick solid line indicates the time evolution of the total dust mass.
\label{fig7}}
\end{figure}

\clearpage

\begin{figure}
\epsscale{1.0}
\plottwo{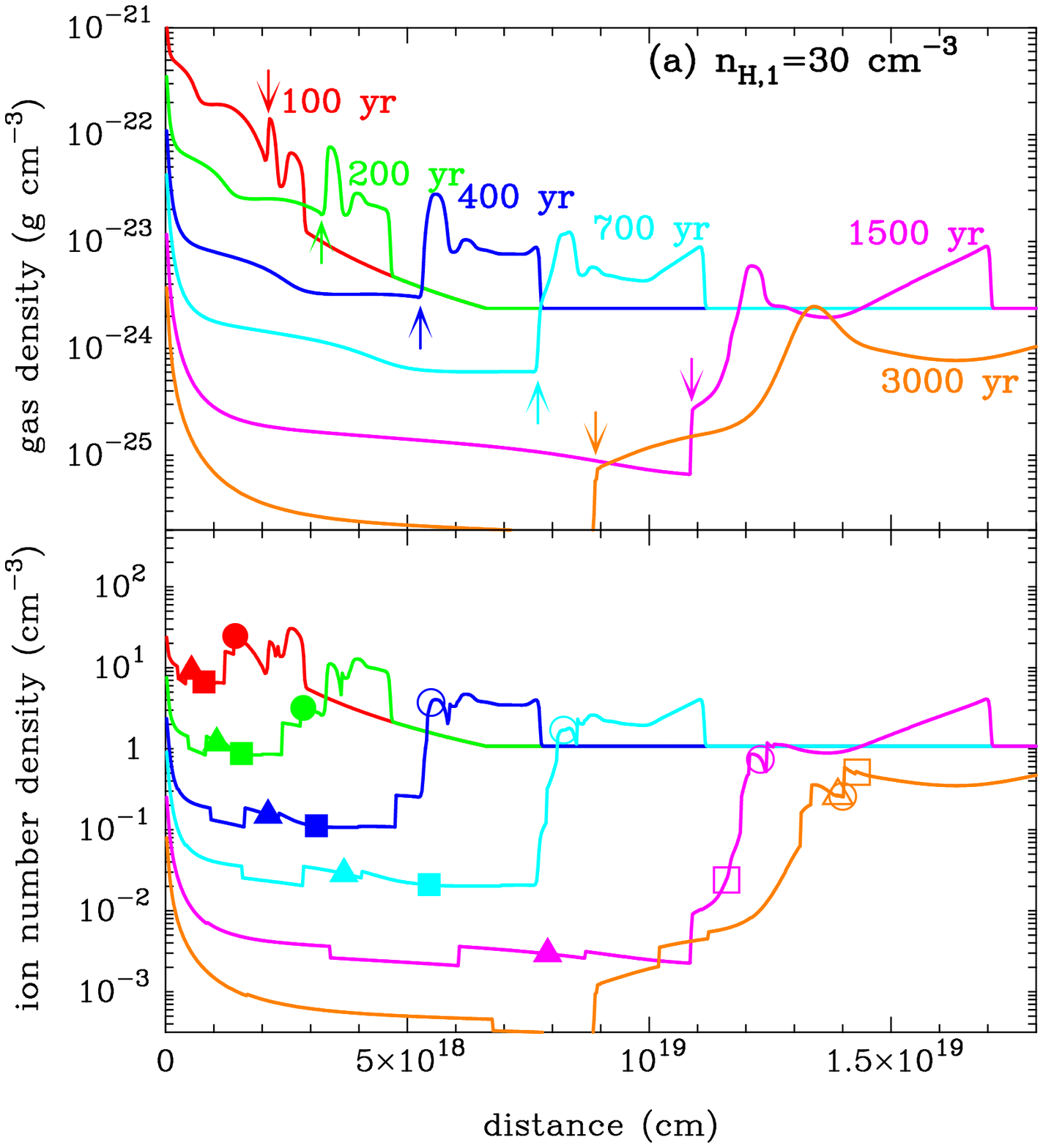}{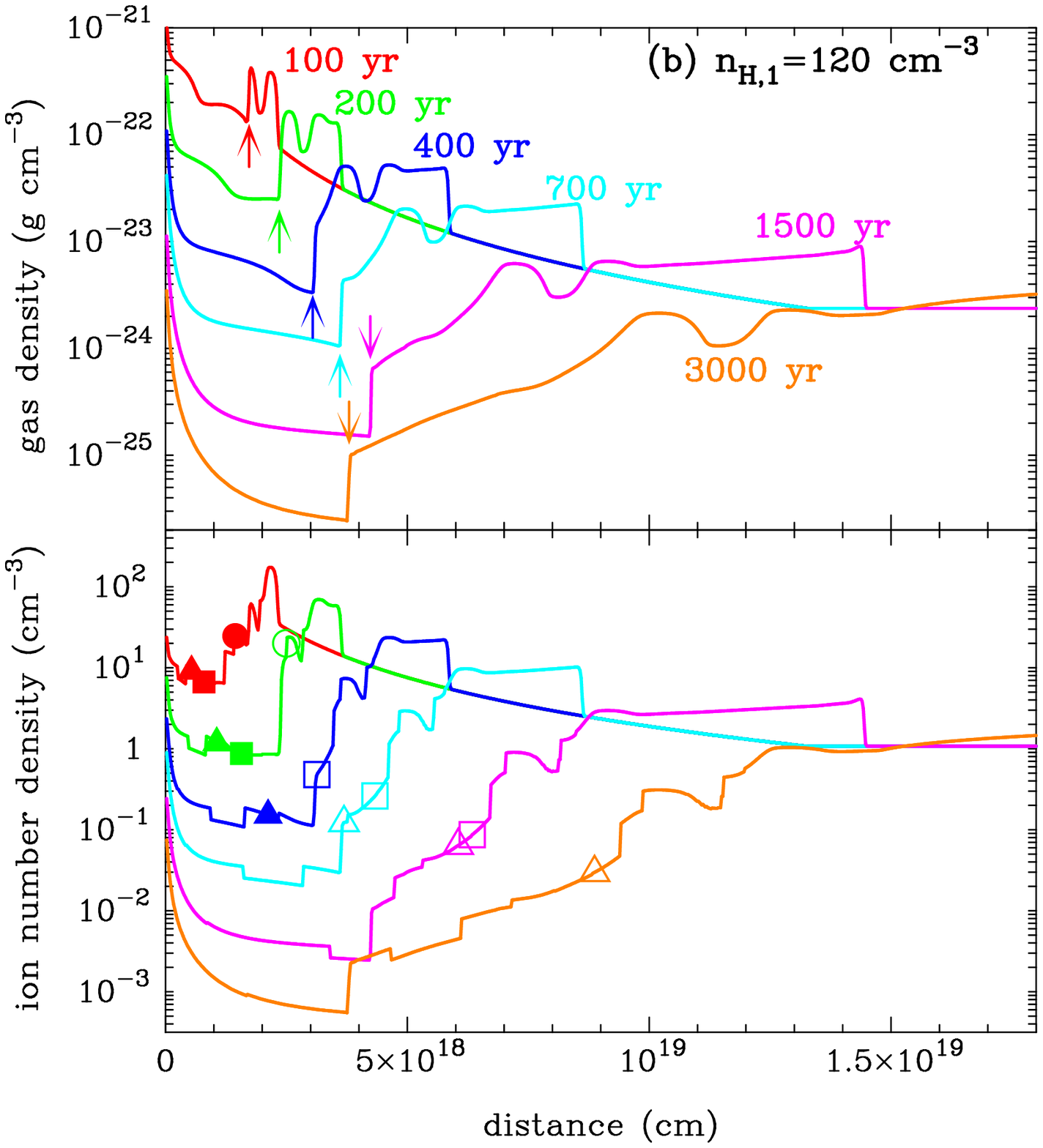}
\caption{ 
 Time evolutions of the gas density ({\it upper panel}) and the 
 corresponding total ion number density ({\it lower panel}) as a function 
 of distance from the center for ({\it a}) $n_{\rm H,1} = 30$ cm$^{-3}$ 
 and ({\it b}) $n_{\rm H,1} = 120$ cm$^{-3}$. 
 The positions of the reverse shock are indicated by the arrows in the 
 upper panels.
 In the lower panels, the positions of C ({\it circles}), SiO$_2$ 
 ({\it squares}), and Si grains ({\it triangles}) are plotted where the 
 filled and opened sympols denote the unshocked and shocked ones, 
 respectively.
\label{fig9}}
\end{figure}

\clearpage

\begin{figure}
\epsscale{1.0}
\plottwo{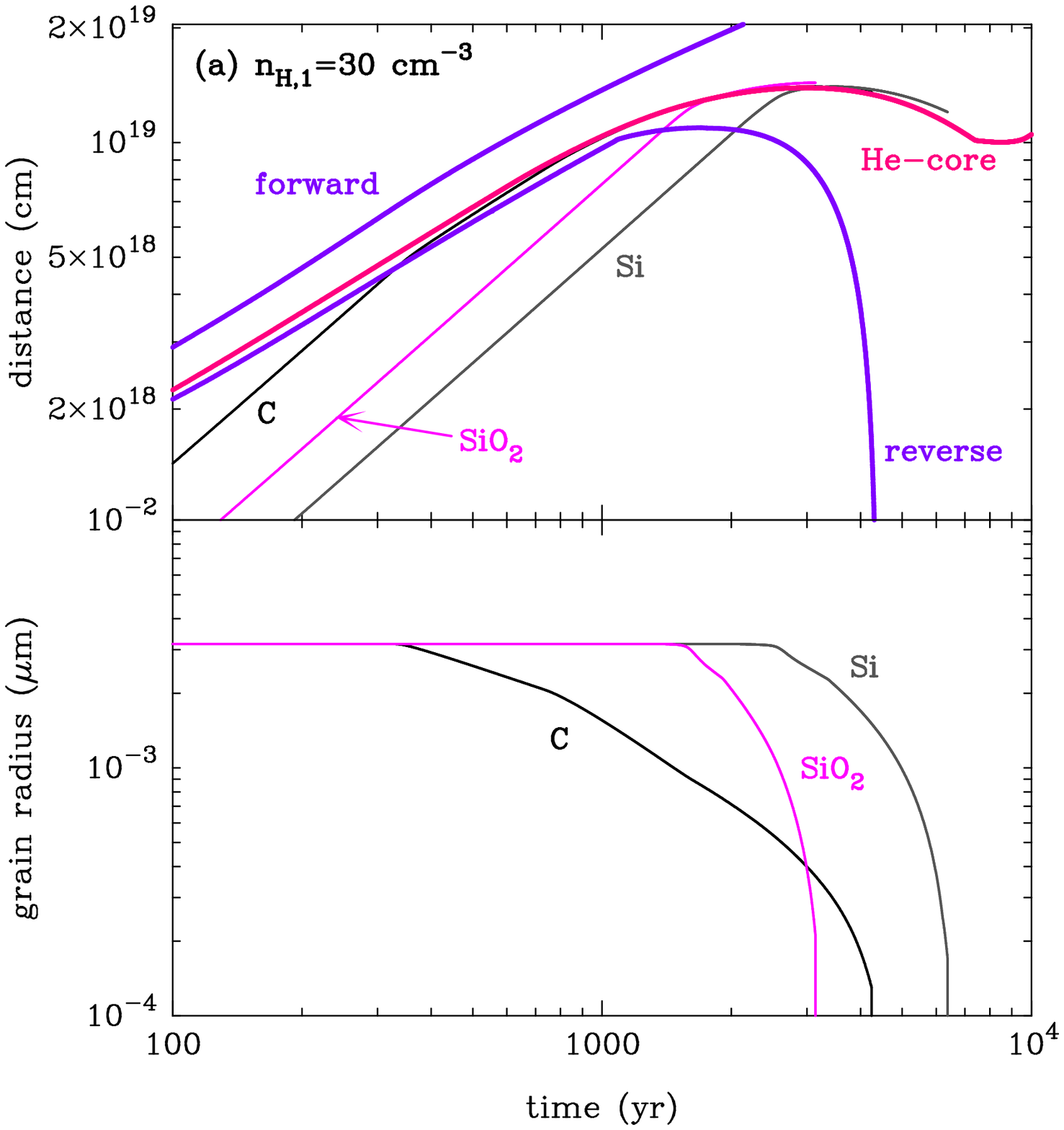}{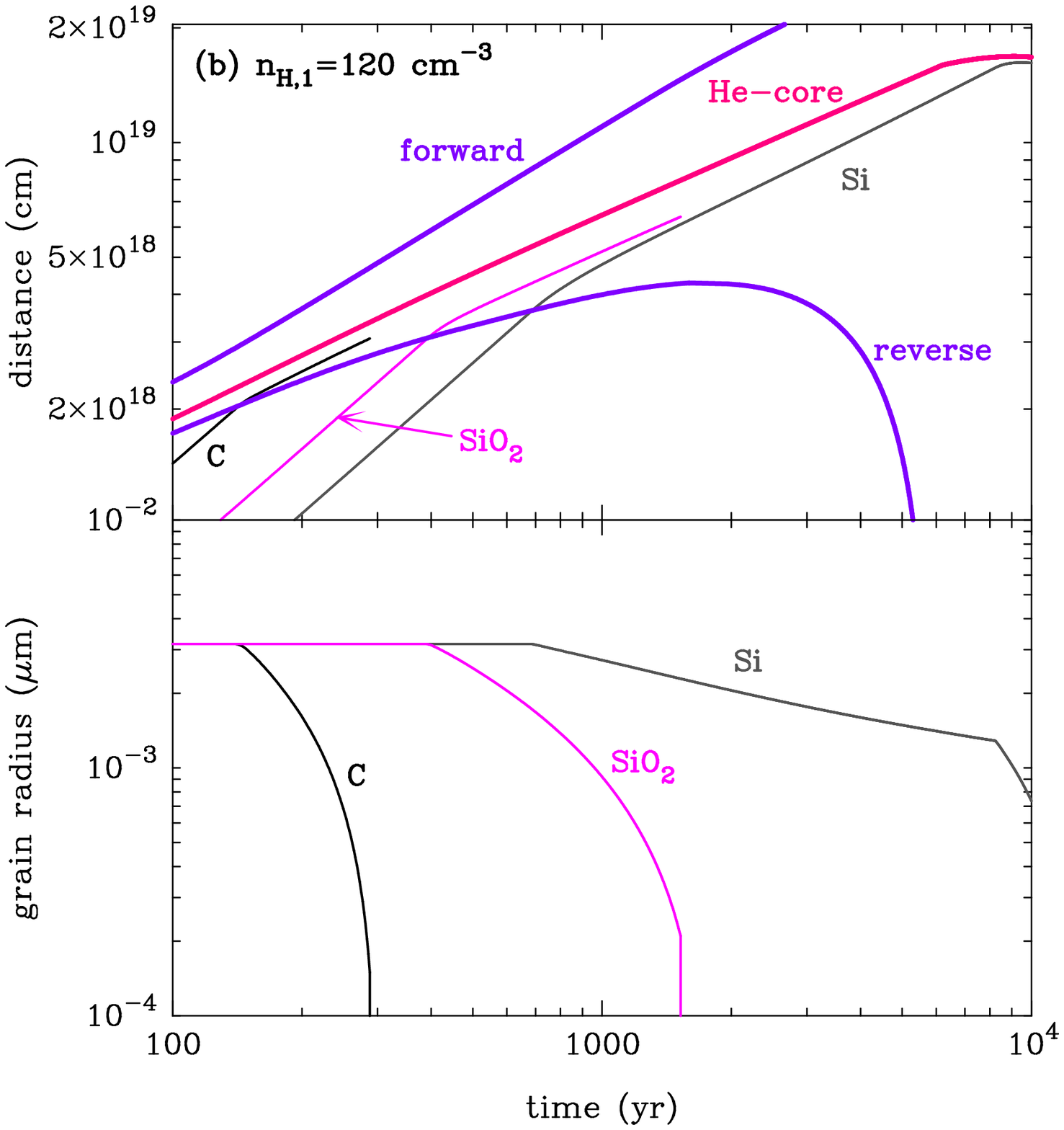}
\caption{ 
 Time evolutions of the positions ({\it upper panel}) and radii ({\it 
 lower panel}) of C, SiO$_2$, and Si grains with the initial radii of 
 0.003 $\mu$m in the Type IIb SNR for ({\it a}) $n_{\rm H,1} = 30$ 
 cm$^{-3}$ and ({\it b}) $n_{\rm H,1} = 120$ cm$^{-3}$.
 The trajectories of the reverse and forward shocks as well as the 
 position of the surface of the He core are depicted by the thick curves 
 in the upper panel.
\label{fig8}}
\end{figure}

\clearpage

\begin{figure}
\epsscale{0.8}
\plotone{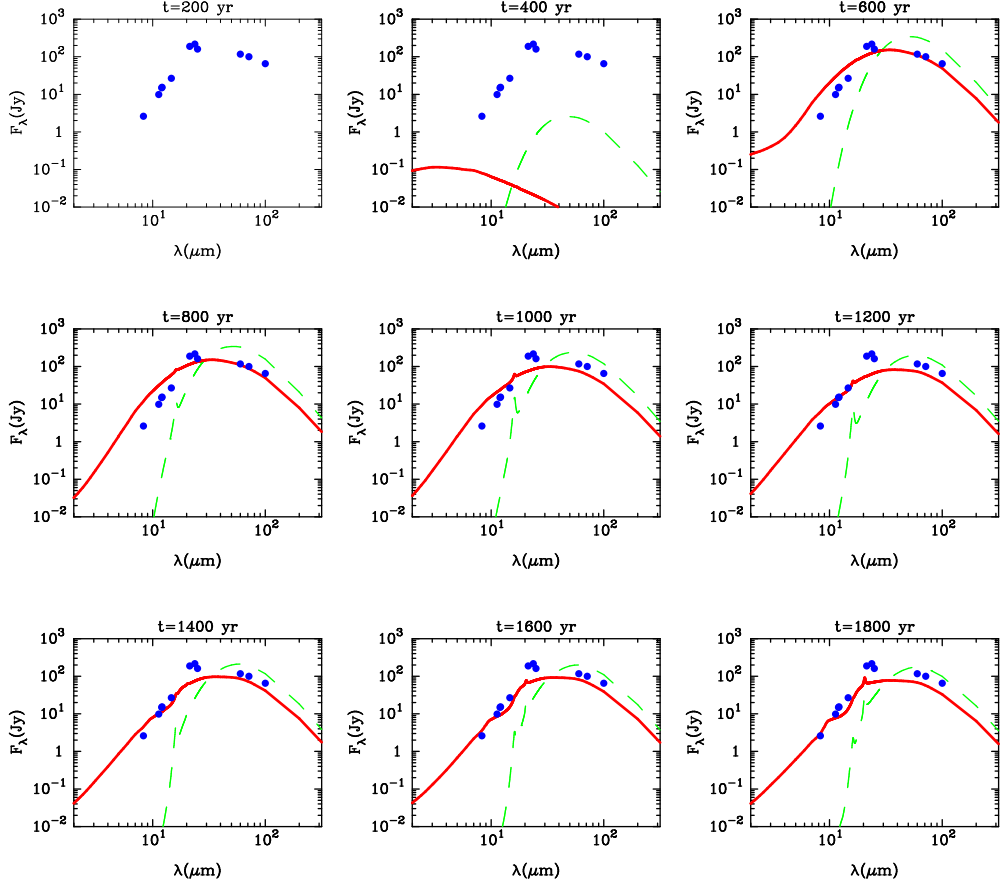}
\caption{ 
 Time evolution of SEDs by thermal emission from shocked-heated dust 
 within the Type IIb SNR for the uniform CSM model A1 
 ($n_{\rm H,0} = 1$ cm$^{-3}$) every 200 yr after the explosion.
 The solid and dashed lines show the IR thermal emission with and without 
 stochastic heating, respectively.
 The data points are the syncrotron-subtracted flux densities observed for 
 Cas A and are taken from Hines et al.\ (2004).
 \label{fig10}}
\end{figure}

\clearpage

\begin{figure}
\epsscale{0.8}
\plotone{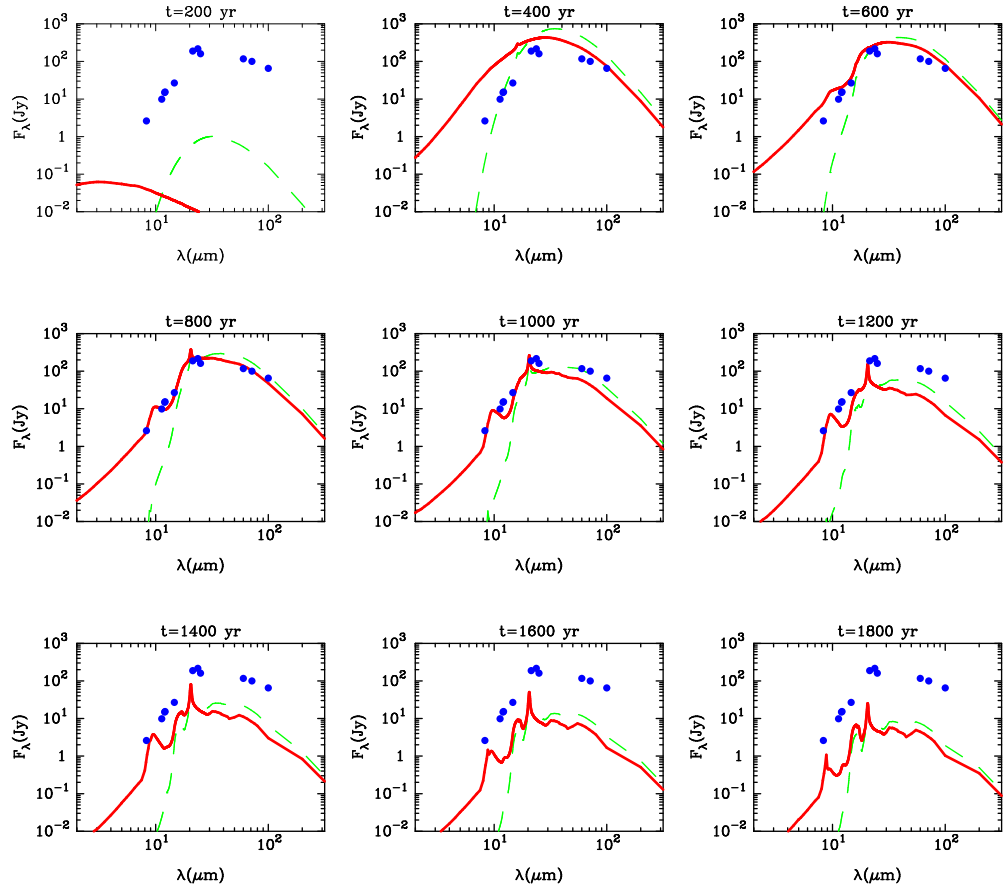}
\caption{ 
 Same as Figure 11, but for the model A10 ($n_{\rm H,0} = 10$ cm$^{-3}$).
 \label{fig11}}
\end{figure}

\clearpage

\begin{figure}
\epsscale{0.8}
\plotone{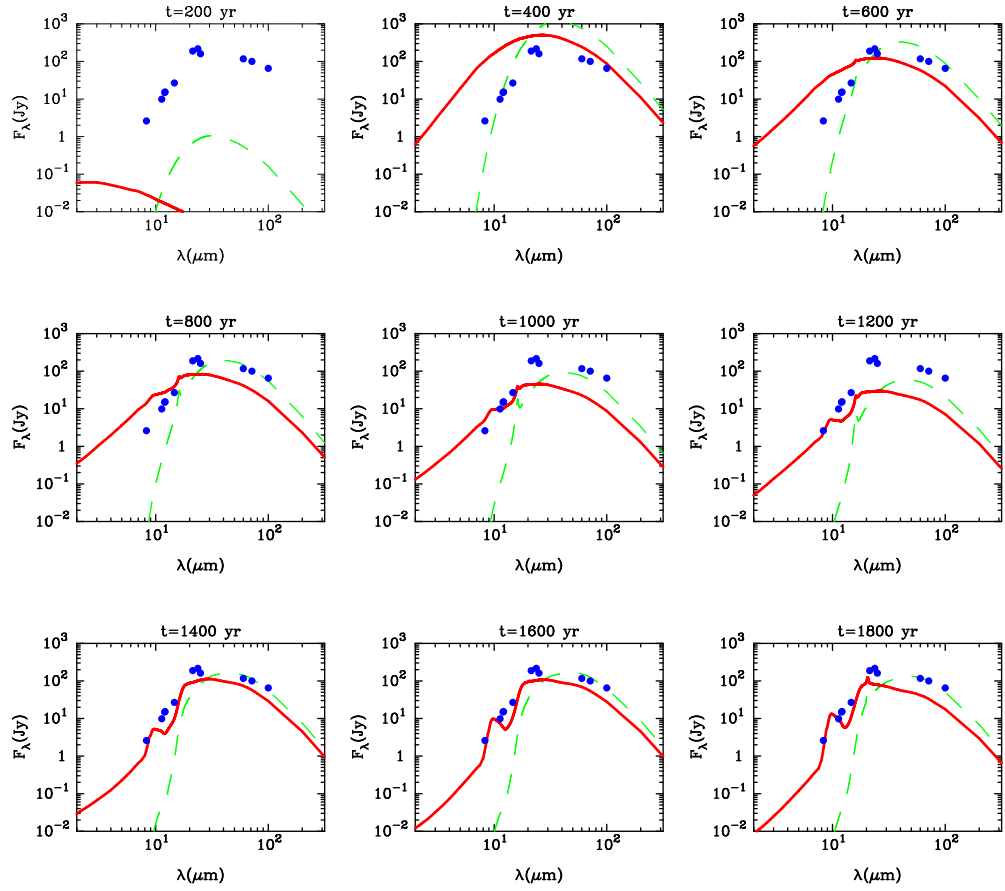}
\caption{ 
 Same as Figure 11, but for the stellar wind CSM model B30 
 ($n_{\rm H,1} = 30$ cm$^{-3}$).
 \label{fig12}}
\end{figure}

\clearpage

\begin{figure}
\epsscale{0.8}
\plotone{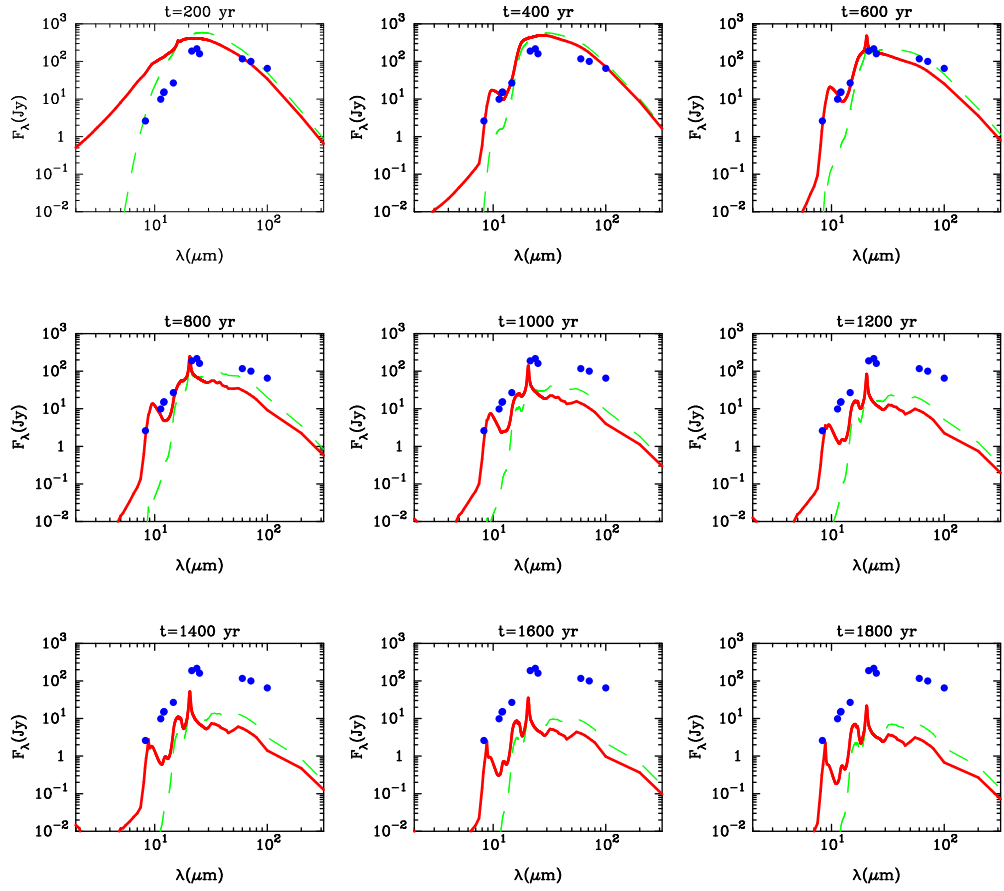}
\caption{ 
 Same as Figure 11, but for the model B120 ($n_{\rm H,1} = 120$ cm$^{-3}$).
 \label{fig13}}
\end{figure}

\clearpage

\begin{figure}
\epsscale{0.5}
\plotone{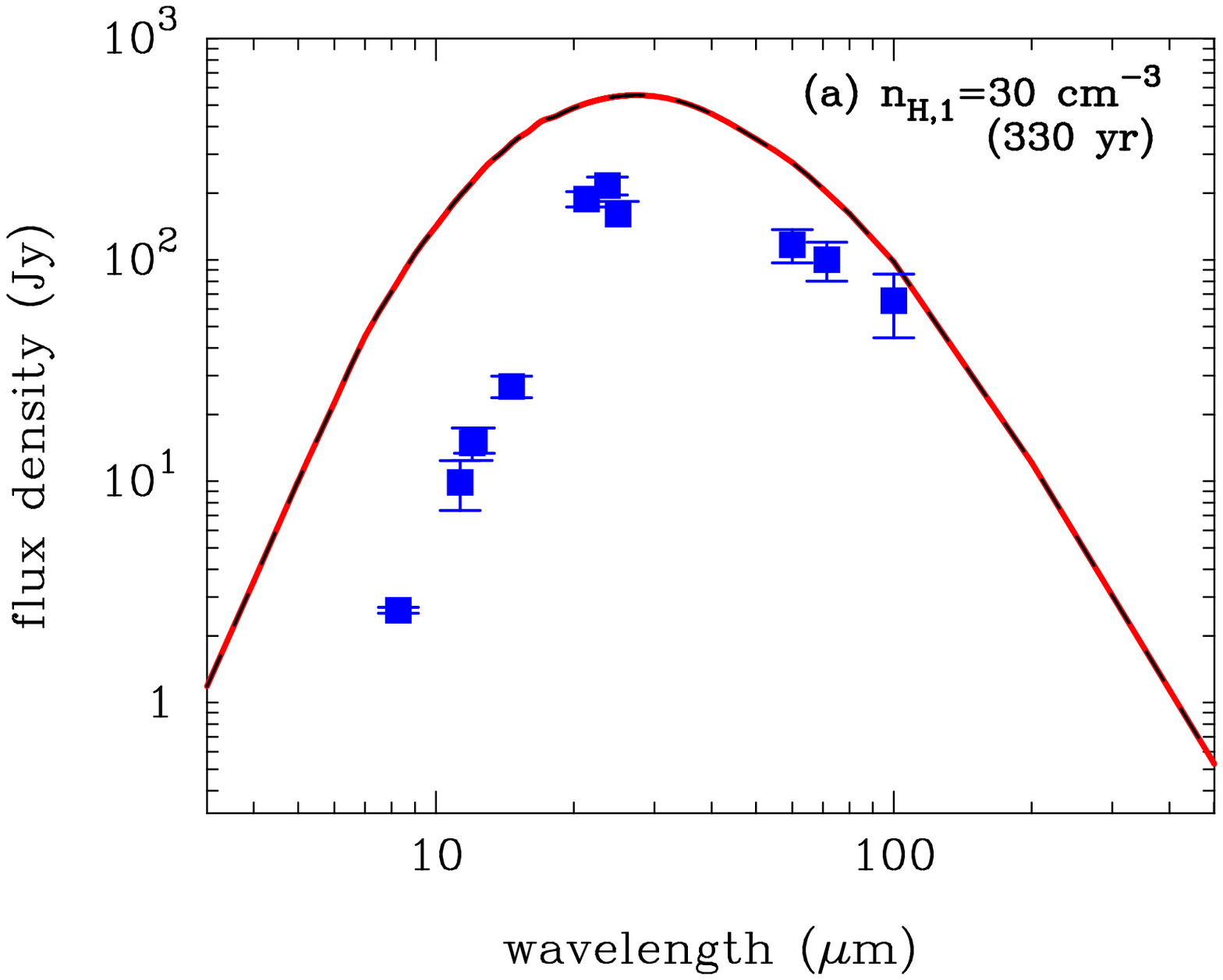}
\plotone{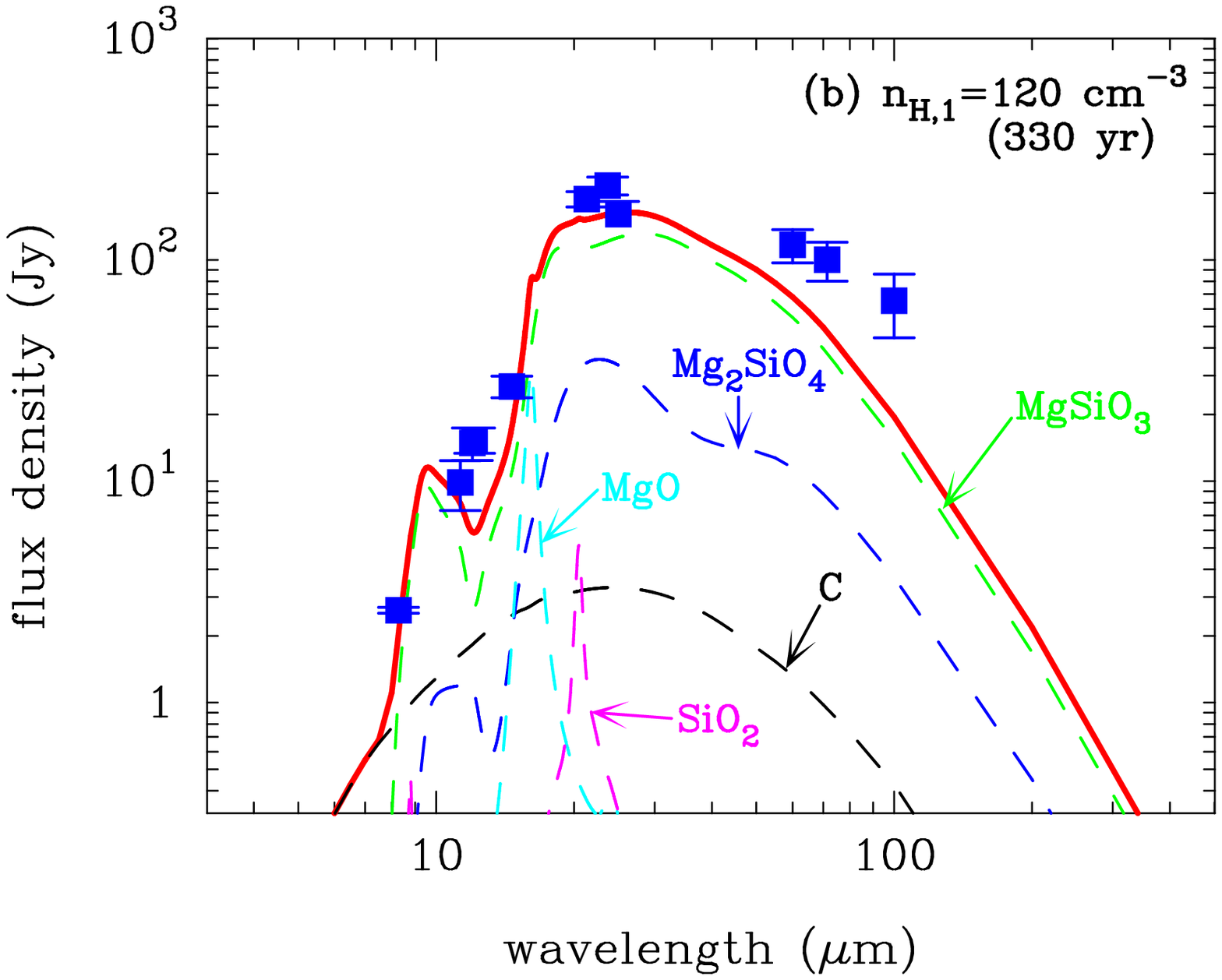}
\plotone{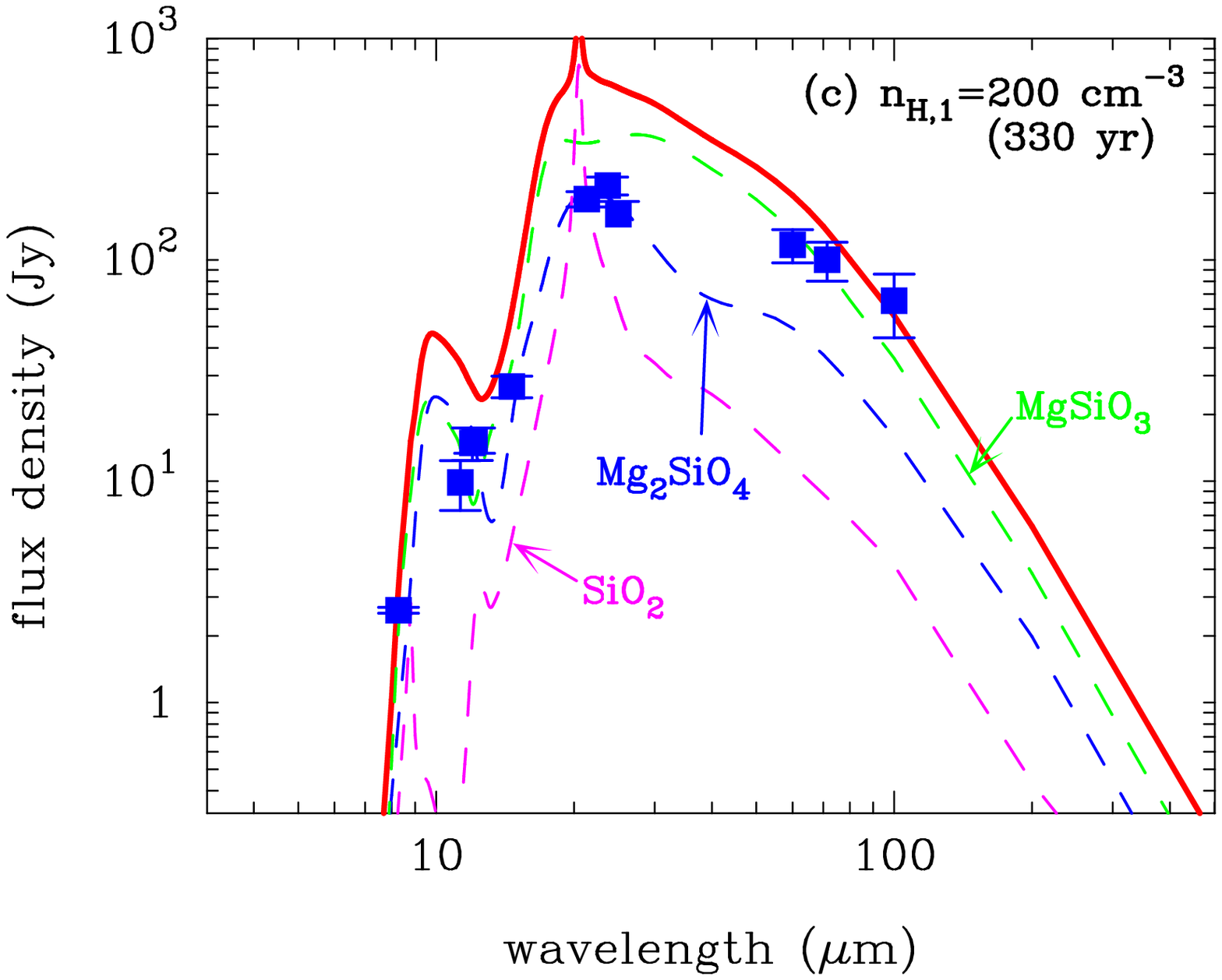}
\caption{
 Comparison between the IR observations of Cas A and the calculated SEDs
 by thermal emission from shocked dust at 330 yr for 
 ({\it a}) $n_{\rm H,1} = 30$ cm$^{-3}$, 
 ({\it b}) 120 cm$^{-3}$, and ({\it c}) 200 cm$^{-3}$.
 The dashed lines indicate the contribution of each dust species.
 The IR flux densities in ({\it a}) are produced only by C grains.
 The data points are the same as in Figure 11.
\label{fig14}}
\end{figure}

\clearpage

\begin{figure}
\epsscale{0.5}
\plotone{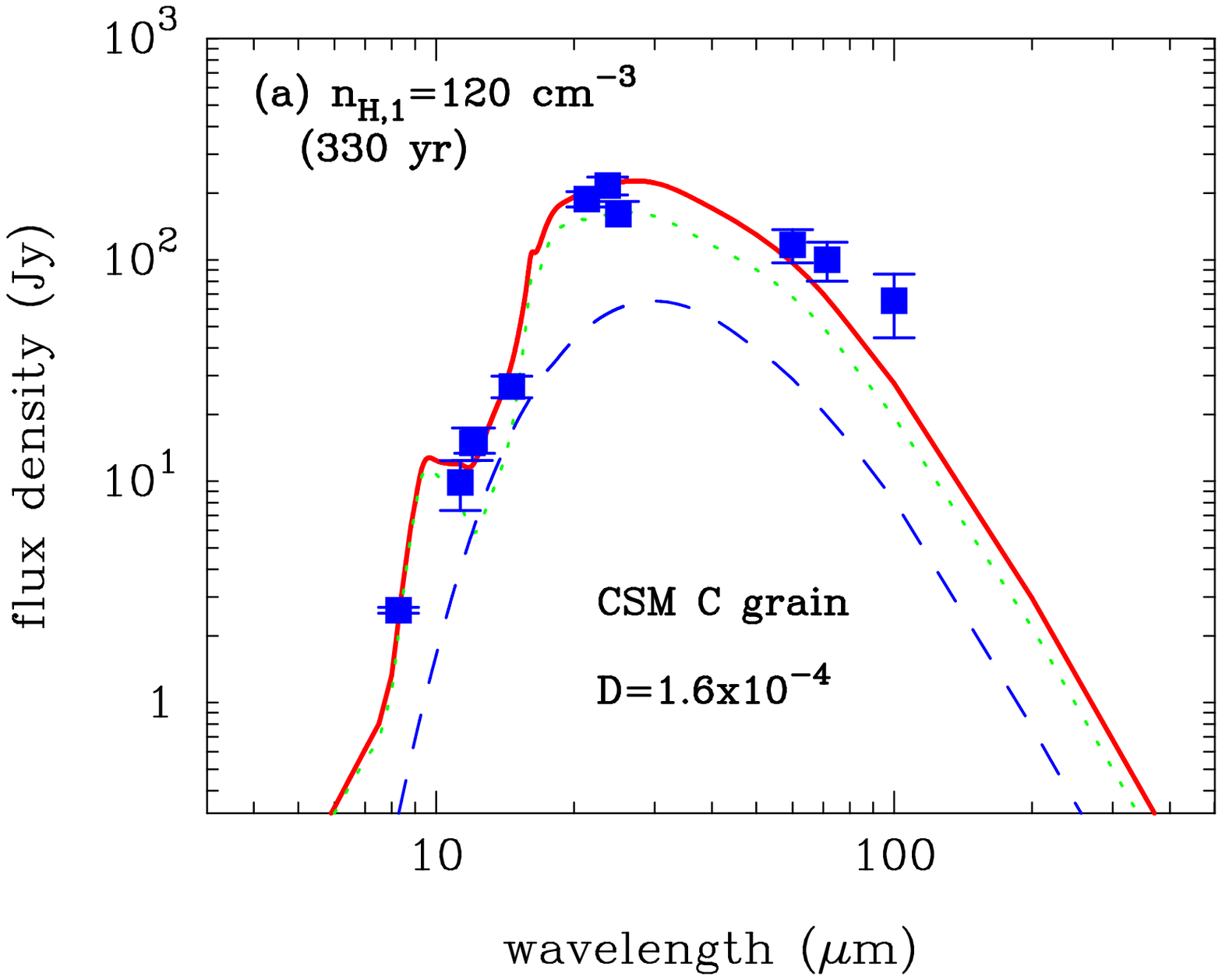}
\plotone{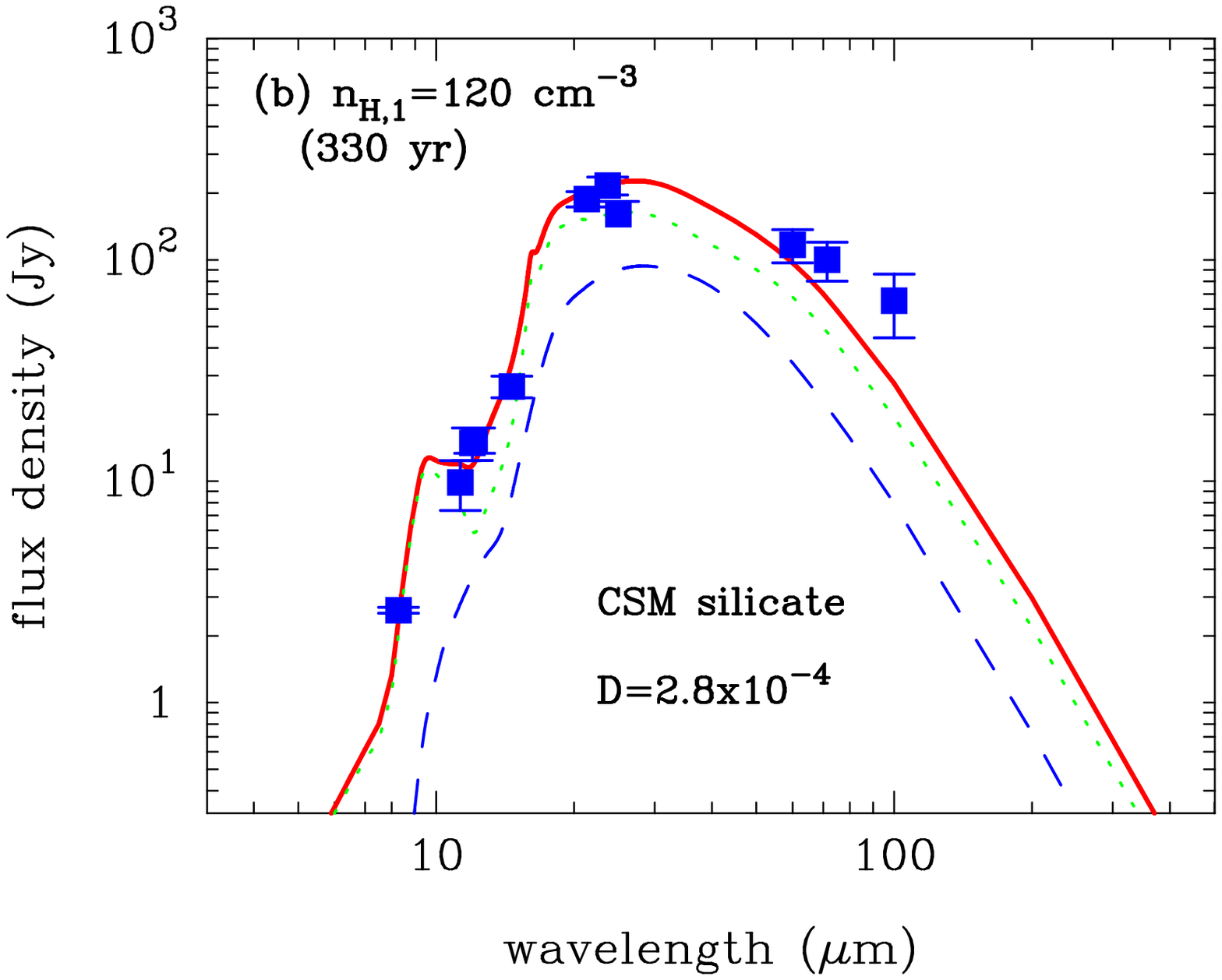}
\caption{
 Contributions from ({\it a}) CS amorphous carbon and ({\it b}) CS 
 astronomical silicate to the IR SED calculated for $n_{\rm H,1} = 120$ 
 cm$^{-3}$.
 The dashed and dotted lines denote the thermal emissions from the 
 shocked CS dust and shocked newly formed dust, respectively, and 
 the solid lines represent the total SED.
 The observation data for Cas A are the same as in Figure 11.
\label{fig15}}
\end{figure}

\clearpage

\begin{figure}
\epsscale{0.5}
\plotone{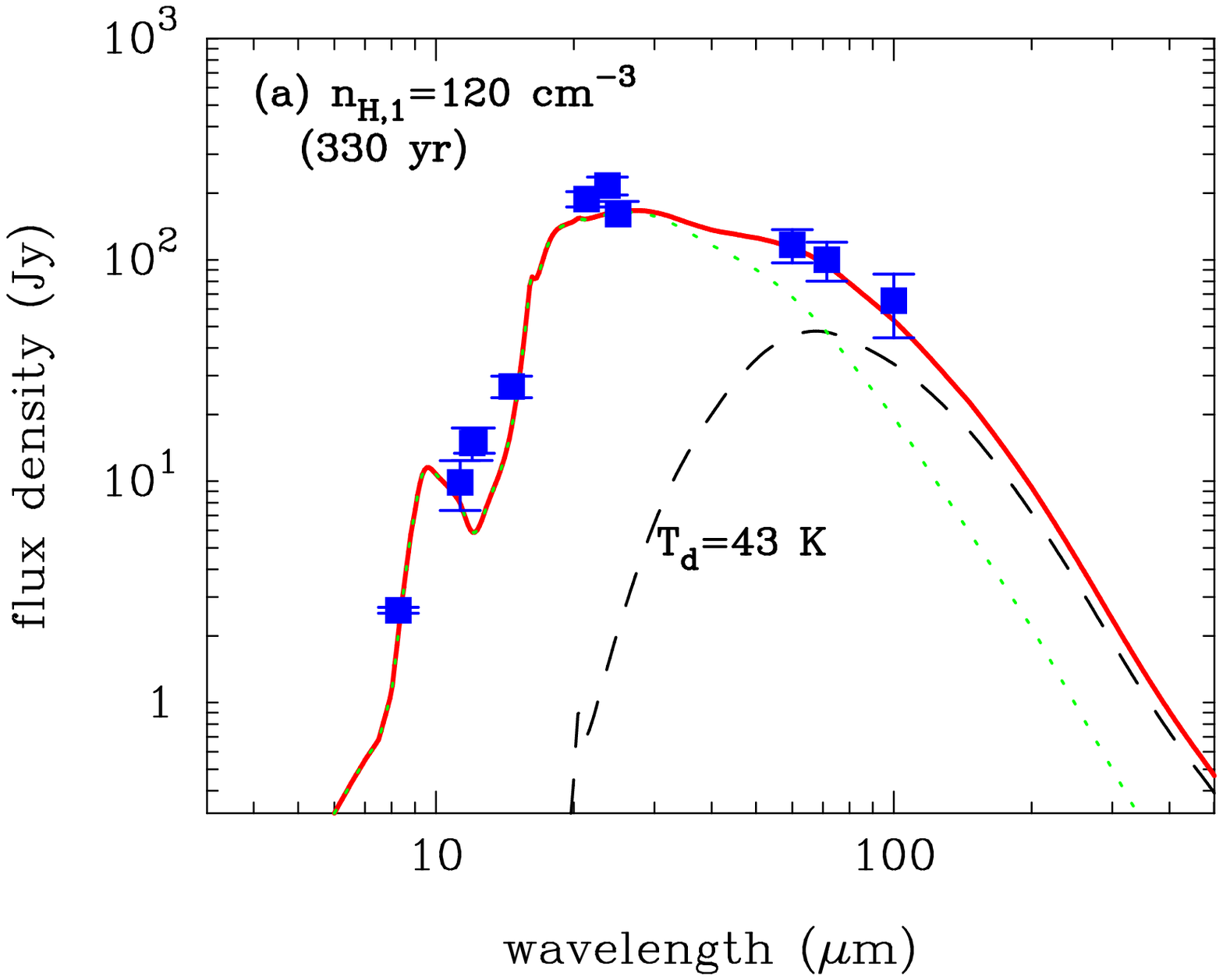}
\plotone{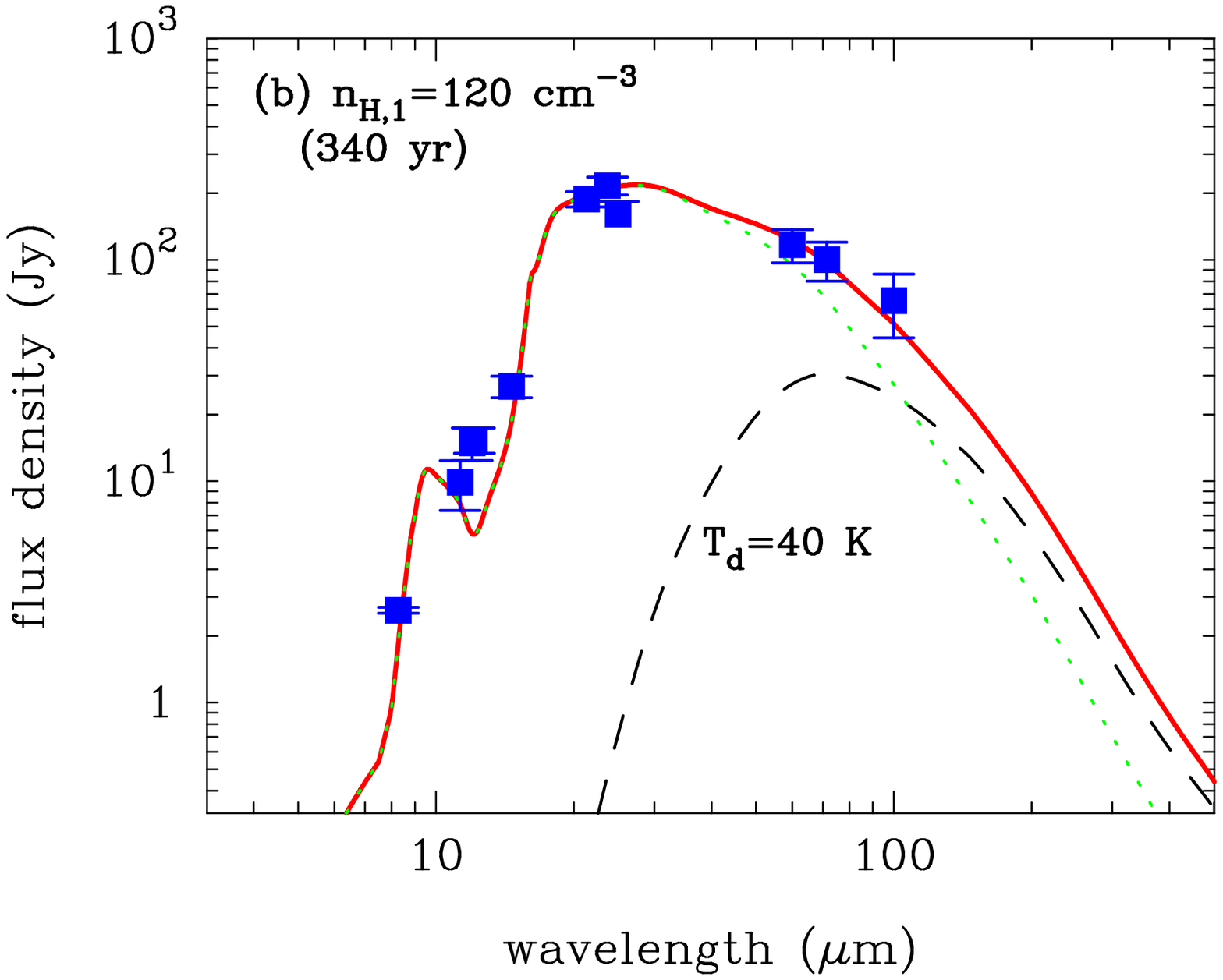}
\caption{
 Contributions from the unshocked ejecta-dust to the IR SED calculated for 
 $n_{\rm H,1} = 120$ cm$^{-3}$ at ({\it a}) 330 yr and ({\it b}) 340 yr. 
 The dashed and dotted lines denote the thermal emissions from the 
 unshocked and shocked newly formed dust, respectively, and the solid 
 lines represent the total SED.
 The temperature of cold dust is 43 K in ({\it a}) and 40 K in ({\it b}).
 The observation data for Cas A are the same as in Figure 11.
\label{fig16}}
\end{figure}

\clearpage

\begin{deluxetable}{lccc}
\tablewidth{0pt}
\tablecaption{Mass of the Main Refractory Species in the Ejecta of 
the SN IIb}
\tablehead{ 
\colhead{species} & 
\colhead{$M_{A, i}$ ($M_\odot$)} &
\colhead{$M_{A, i}^{\rm dust}$ ($M_\odot$)} & 
\colhead{$M_{A, i}^{\rm dust}/M_{A, i}$} \\
\colhead{(1)} & \colhead{(2)} &
\colhead{(3)} & \colhead{(4)}
}
\startdata
C      & $1.14 \times 10^{-1}$ & $7.08 \times 10^{-2}$ & 0.621   \\
O      & $6.86 \times 10^{-1}$ & $4.35 \times 10^{-2}$ & 0.063   \\
Mg     & $1.07 \times 10^{-1}$ & $2.05 \times 10^{-2}$ & 0.192   \\
Al     & $9.31 \times 10^{-3}$ & $3.28 \times 10^{-5}$ & 0.004   \\
Si     & $1.07 \times 10^{-1}$ & $3.12 \times 10^{-2}$ & 0.292   \\
S      & $3.33 \times 10^{-2}$ & $5.35 \times 10^{-4}$ & 0.016   \\
Fe     & $7.92 \times 10^{-2}$ & $9.35 \times 10^{-4}$ & 0.012   \\
others & $1.61 \times 10^{-1}$ & 0 & 0   \\ \hline
Total  & 1.30 & $0.167$ & ---
\enddata
\tablecomments{
Col.\ (1): Species of the main heavy element in the ejecta of the SN IIb.
Col.\ (2): Mass of each heavy element ($M_{A, i}$).
Col.\ (3): Mass of each element that is locked up in dust grains at the 
time of dust formation ($M_{A, i}^{\rm dust}$).
Col.\ (4): Mass fraction of each element that is locked up in dust grains
at the time of dust formation ($M_{A, i}^{\rm dust}/M_{A, i}$).
}
\end{deluxetable}

\clearpage

\begin{deluxetable}{lccc}
\tablewidth{0pt}
\tablecaption{Mass of Each Dust Species Formed in the Ejecta of the 
SN IIb}
\tablehead{ 
\colhead{dust species} & \colhead{$M_{d, j}$ ($M_\odot$)} &
\colhead{$M_{d, j}/M_d$} & \colhead{$M_{d, j}/M_A$}  \\
\colhead{(1)} & \colhead{(2)} &
\colhead{(3)} & \colhead{(4)}
}
\startdata
C             & $7.08 \times 10^{-2}$ & $0.423$ & 0.054   \\
Al$_2$O$_3$   & $6.19 \times 10^{-5}$ & $3.7 \times 10^{-4}$ 
              & $4.8 \times 10^{-4}$   \\
MgSiO$_3$     & $5.46 \times 10^{-2}$ & $0.326$ & 0.042   \\
Mg$_2$SiO$_4$ & $1.74 \times 10^{-2}$ & $0.104$ & 0.013   \\
SiO$_2$       & $1.57 \times 10^{-2}$ & $0.094$ & 0.012   \\
MgO           & $2.36 \times 10^{-3}$ & $0.014$ & $1.8 \times 10^{-3}$   \\
FeS           & $1.47 \times 10^{-3}$ & $0.009$ & $1.1 \times 10^{-3}$   \\
Si            & $5.07 \times 10^{-3}$ & $0.030$ & $3.9 \times 10^{-3}$   \\ \hline
Total         & 0.167 & $1$ & 0.128 
\enddata
\tablecomments{
Col.\ (1): Species of dust formed in the ejecta of the SN IIb.
Col.\ (2): Mass of each grain species ($M_{d, j}$).
Col.\ (3): Ratio of mass of each dust species to the total dust mass 
($M_{d, j}/M_d$).
Col.\ (4): Ratio of mass of each dust species to the total metal mass 
($M_{d, j}/M_A$).}

\end{deluxetable}

\clearpage

\begin{deluxetable}{lccccccccc}
\tablewidth{0pt}
\tablecaption{Results of the dust evolution calculation for each CSM model}
\tablehead{ 
\colhead{CSM model} & 
\colhead{$t_{\rm coll}^{\rm He}$} & 
\colhead{$t_{\rm coll}^{\rm O}$} & 
\colhead{$t_{\rm coll}^{\rm Si}$} & 
\colhead{$M_{\rm surv}$} &
\colhead{$\eta$} &
\colhead{$R_{\rm rev}^{330}$} &
\colhead{$R_{\rm forw}^{330}$} &
\colhead{$M_{\rm warm}^{330}$} &
\colhead{$M_{\rm cool}^{330}$}  \\ 
\colhead{} & 
\colhead{(yr)} & 
\colhead{(yr)} & 
\colhead{(yr)} & 
\colhead{($M_\odot$)} &
\colhead{($M_{\rm surv}/M_{\rm d}$)} &
\colhead{(pc)} & 
\colhead{(pc)} & 
\colhead{($M_\odot$)} & 
\colhead{($M_\odot$)} \\
\colhead{(1)} & \colhead{(2)} & \colhead{(3)} & \colhead{(4)} &
\colhead{(5)} & \colhead{(6)} & \colhead{(7)} & \colhead{(8)} &
 \colhead{(9)} & \colhead{(10)}}
\startdata
A0.1   & 335   & 1300  & 3400 &
$1.3 \times 10^{-4}$   & $7.8 \times 10^{-4}$  & 
2.76   & 3.46  & 0.0   & 0.167    \\
A1     & 155   & 590   & 1550 &
0.0            & 0.0          &
1.98   & 2.52  & $1.0 \times 10^{-4}$ & 0.167     \\
A10    & 75    & 270   & 700  &
0.0            & 0.0          &
1.32   & 1.83  & 0.038 & 0.094    \\
\hline
B30    & 63    & 340   & 1850 &
0.0            & 0.0          & 
1.51   & 2.20  & 0.048 & 0.111    \\
B120   & 48    & 150   & 560  &
0.0            & 0.0          & 
0.93   & 1.67  & 0.008 & 0.072    \\
B200   & 45    & 120   & 370  &
0.0            & 0.0          & 
0.70  & 1.50  & 0.028 & 0.011     \\
\enddata

\tablecomments{
Col.\ (1): Name of the CSM models considered in this paper.
The labels A and B represent the uniform CSM and the stellar wind CSM 
($\rho \propto r^{-2}$), respectively, and the attached number denotes 
the value of $n_{\rm H,0}$ for model A and $n_{\rm H,1}$ model B 
(see text).
Cols.\ (2), (3), and (4): 
Times at which the reverse shock encounters the He core, O-rich layer, and
Si-rich layer, respectively.
Col.\ (5): Mass of surviving dust at $t=10^6$ yr.
Col.\ (6): Ratio of surviving dust mass $M_{\rm surv}$ to the initial 
           dust mass $M_d$.
Cols.\ (7) and (8): Radii of the reverse and forward shocks at $t =$ 330 yr.
Col.\ (9): Mass of warm dust heated up by the shock at $t =$ 330 yr.
Col.\ (10): Mass of unshocked cold dust at $t =$ 330 yr.
}

\end{deluxetable}

\end{document}